\documentclass[acmsmall,screen,nonacm,anonymous=false]{acmart}

\acmJournal{TOMPECS}

\usepackage{tikz}
\usetikzlibrary{positioning,fit,calc,arrows.meta,backgrounds,patterns,decorations.pathreplacing}
\usepackage{enumitem}

\definecolor{vllm}{HTML}{1A2B4C}     %
\definecolor{sglang}{HTML}{A8432F}   %

\tikzset{
  paperfig/.style   = {font=\footnotesize, line join=round, line cap=round,
                       >={Latex[length=1.7mm,width=1.3mm]}},
  comp/.style       = {draw=black!80, fill=white, line width=0.5pt,
                       rounded corners=1pt, align=center, inner sep=3pt},
  compshade/.style  = {comp, fill=gray!10},
  setting/.style    = {comp, dashed},
  region/.style     = {draw=black!70, fill=gray!6, line width=0.5pt,
                       rounded corners=2pt},
  regionstrong/.style = {region, line width=1.2pt},   %
  cluster/.style    = {draw=black!60, dashed, line width=0.5pt,
                       rounded corners=3pt},
  title/.style      = {font=\footnotesize\bfseries, text=black},
  lbl/.style        = {font=\scriptsize, text=black!85},
  note/.style       = {font=\scriptsize\itshape, text=black!70},
  num/.style        = {circle, draw=black!85, fill=white, line width=0.5pt,
                       inner sep=0pt, minimum size=3.4mm,
                       font=\tiny\bfseries},
  reqpath/.style    = {->, draw=black, line width=1.0pt},
  telemetry/.style  = {->, draw=black!80, line width=0.6pt, densely dotted},
  offline/.style    = {->, draw=black!70, line width=0.6pt, dashed},
  plain/.style      = {->, draw=black!80, line width=0.6pt},
  token/.style      = {draw=black, fill=black!75, line width=0.3pt},
  prefill/.style    = {draw=black!80, line width=0.5pt,
                       pattern=north east lines, pattern color=black!55},
  iterline/.style   = {draw=black!25, line width=0.4pt},
}

\begin{document}

\title{A Controlled Experimental Evaluation of Configuration and Cache
       Architecture in LLM Serving Engines}

\author{Muhammad Faizan Anwar}
\affiliation{%
  \institution{Thompson Rivers University}
  \city{Kamloops}
  \state{British Columbia}
  \country{Canada}}
\email{T00804579@mytru.ca}

\author{Sina Keshvadi}
\affiliation{%
  \institution{Thompson Rivers University}
  \city{Kamloops}
  \state{British Columbia}
  \country{Canada}}
\email{skeshvadi@tru.ca}

\renewcommand{\shortauthors}{Anwar and Keshvadi}

\begin{abstract}
Multi-agent LLM applications repeat most of their prompt tokens across
requests, and serving engines reuse the cached state of shared prefixes. vLLM
and SGLang reuse it at different granularities, and benchmarks of their
default deployments report large performance gaps, usually credited to the
cache. Such comparisons rarely control for other configuration differences.
One of them, whether a long prompt is prefilled in chunks, is set by each
engine's defaults from the hardware and the release, and it confounds the
comparison. We built a reproducible testbed on Kubernetes and a seeded
multi-agent workload generator, and compared the two engines under vendor
defaults and with chunked prefill matched. SGLang's configuration is the same
in both arms, which makes it an internal control across independently
provisioned deployments. Under defaults, vLLM's median time to first token is
up to $64\times$ SGLang's. Matching one setting removes 78\% of the capacity
gap where that gap is largest and 30 to 67\% of the latency gap. Near
saturation the remaining latency gap reflects queue growth. A fluid queue
model driven only by a 4.4\% throughput shortfall reproduces it, and shows
that such a gap scales with the length of the benchmark run. What survives
matching is a tail-latency advantage of 2.4 to $6.4\times$ at equal
throughput, consistent with cache-aware ordering of waiting requests. We
derive six recommendations for comparing serving engines.
\end{abstract}


\maketitle

\section{Introduction}
\label{sec:intro}

Applications built on large language models increasingly run as multi-agent
workflows. A planner decomposes a task, workers call tools and reason over
retrieved documents, and an aggregator merges their answers. Every call in
such a workflow repeats the same system instructions, tool schemas and shared
context. Across five agent and long-context workloads, 85 to 97\% of prompt
tokens are shared with at least one other request~\cite{srivatsa2025preble},
and in a production LLM-based search service over 94\% of request tokens
recur across users~\cite{lin2024parrot}. Recomputing that shared state wastes
the resource that limits a serving GPU. The limit is memory, not arithmetic
(Section~\ref{sec:background}).

Two widely used open-source serving engines reuse shared prompts in different
ways. vLLM stores its key-value (KV) cache in fixed-size blocks and reuses a
block when its contents match a cached one~\cite{kwon2023vllm}. SGLang keeps
cached tokens in a radix tree, RadixAttention, and reuses a shared prefix of
any length~\cite{zheng2024sglang}. Operators choose between such engines, and
size GPU fleets around them, on the evidence of benchmarks. The SGLang authors
report throughput up to $6.4\times$ higher than contemporaneous
systems~\cite{zheng2024sglang}, and several serving systems for agent
workloads take finer-grained prefix reuse as their starting
point~\cite{pan2025kvflow,bian2025tokencake,srivatsa2025preble}.

We deployed both engines on one NVIDIA L4 GPU under their documented defaults
and measured the gap such benchmarks lead one to expect. It is large. With 8,192-token
prompts that share 90\% of their tokens, vLLM's median time to first token
(TTFT) is $64.1\times$ SGLang's. Near vLLM's saturation point the median gap
is $36.4\times$, and the highest throughput SGLang sustains is 1.94 requests
per second against 1.40 for vLLM. A gap of this size invites a simple
explanation: the finer-grained cache wins.

That explanation has not been tested. The engines' defaults differ in a
second respect that has nothing to do with caching. A long prompt's prefill
can run as one uninterrupted step, during which every running request
waits, or it can be split into chunks interleaved with ongoing
decoding~\cite{agrawal2024sarathi}. This is a scheduler setting. On our
hardware SGLang chunks long prefills by default and vLLM does not. Neither
engine is misconfigured: vLLM 0.6.3 enables chunked prefill only above a
32,768-token context, and SGLang 0.4.1 enables it on any GPU with less than
25~GB of memory. A comparison of default deployments therefore varies two
factors at once. It cannot say which of them produced the gap.

The particular pairing has since changed upstream. vLLM's V1 engine enables
chunked prefill by default wherever possible, where V0 enabled it
conditionally~\cite{vllm2025v1guide}. The confound is a property of defaults
rather than of one release. Defaults depend on the GPU and the configured
context length, and they change between versions, so a benchmark that does
not record the configuration each engine actually resolved cannot state which
pairing it measured. Comparisons made against V0-era vLLM, including the
baselines behind widely cited speedups, fall within the configuration space
we study.

This paper separates the two factors in a controlled experiment and asks four
questions:
\begin{itemize}[leftmargin=2.6em,itemsep=2pt,topsep=4pt]
  \item[\textbf{RQ1}] How large is the performance gap between the two engines
        under their default configurations, and where in the load range does
        it appear?
  \item[\textbf{RQ2}] How much of that gap remains once both engines use the
        same chunked-prefill configuration?
  \item[\textbf{RQ3}] Why do latency and capacity measures of the remaining
        gap disagree near saturation?
  \item[\textbf{RQ4}] Does the gap grow with the degree of prefix sharing, as
        an explanation based on cache architecture predicts?
\end{itemize}

We built a testbed on Google Kubernetes Engine and a workload generator that
reproduces bursty arrivals, agent dependency graphs and exact prefix sharing
from a seed. We ran each engine under its vendor defaults and again with
chunked prefill matched at a 2,048-token budget. Only vLLM's setting
changes between the two arms. SGLang resolves the same
configuration in both arms, so it serves as an internal control. Rebuilt on
an independently provisioned cluster days later, its latencies agree with the
first deployment within 0.5 to 11.7\%. We report 99 runs and 63,288 requests,
analysed as comparisons paired by seed.

\begin{figure}[t]
  \centering
  \includegraphics{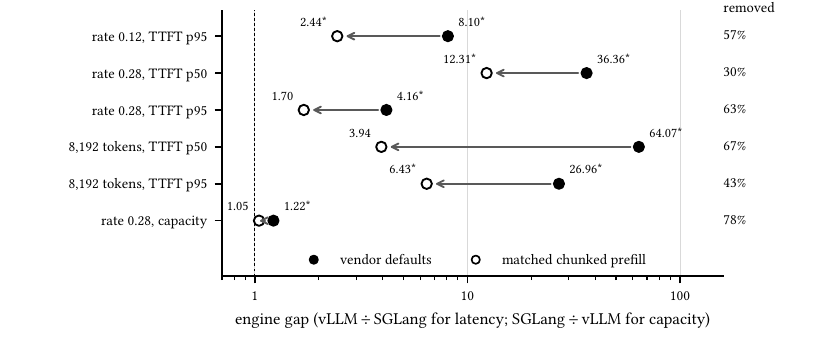}
  \caption{One setting accounts for most of the gap. Each row shows the engine gap under vendor defaults (filled) and with chunked
    prefill matched on both engines (hollow), for every condition measured in
    both configurations. Latency gaps are vLLM's TTFT divided by SGLang's; the
    capacity gap is SGLang's sustained throughput divided by vLLM's. The right
    column gives the fraction of each default gap, on a log scale, that
    matching removes. Geometric means of three seed-paired runs; $^*$ marks a
    paired $t$-test on log values with $p<0.05$.}
  \label{fig:decomp}
\end{figure}

Figure~\ref{fig:decomp} summarises the answer. Matching one setting removes
78\% of the capacity gap at the load where that gap is largest, and 30 to
67\% of the latency gap across conditions. At that load the matched engines
differ in throughput by 4.6\%, yet their median TTFT still differs by
$12.3\times$. The small shortfall leaves vLLM just short of the offered load,
and beyond that point TTFT measures the growth of a queue rather than the
speed of an engine. A fluid queue model driven only by the throughput
shortfall reproduces the effect, and shows that the gap grows with the length
of the benchmark run. What survives matching is narrower. At equal throughput
SGLang retains a $2.4$ to $6.4\times$ advantage at the 95th percentile of
TTFT, consistent with its cache-aware ordering of waiting requests but not
isolated to it. The gap also fails the prediction of RQ4: it narrows as
sharing grows and reverses at high sharing, a pattern the scheduling
difference predicts as well as the cache does.

This paper makes four contributions:
\begin{enumerate}[leftmargin=1.6em,itemsep=2pt,topsep=4pt]
  \item \textbf{A controlled decomposition.} We separate the performance gap
        between vLLM and SGLang into the part due to chunked-prefill configuration
        and the part that remains, with one engine serving as an internal
        control
        (Sections~\ref{sec:design} and~\ref{sec:results}).
  \item \textbf{A methodological result.} Near saturation, TTFT amplifies small differences in capacity nonlinearly, so latency must be
        reported together with sustained throughput
        (Sections~\ref{sec:metrics} and~\ref{sec:results}).
  \item \textbf{A located residual.} We characterise what survives matching through per-request
        distributions, tails and run-to-run variability
        (Section~\ref{sec:variability}).
  \item \textbf{A reproducible apparatus.} We release a testbed defined entirely in code, a
        seeded workload generator, and the per-run logs behind every number
        reported here (Section~\ref{sec:system} and
        Appendix~\ref{app:artifact}).
\end{enumerate}

\section{Background}
\label{sec:background}

\subsection{Prefill, decode and the KV cache}
\label{sec:bg-phases}

Serving a request takes two phases. \emph{Prefill} processes every prompt
token in parallel and produces the first output token; it is a large matrix
computation that keeps the GPU's arithmetic units busy. \emph{Decode} then
produces one token per step. Each step reads the model's weights and the
request's cached state to compute a single new token, so decode is limited by
memory bandwidth rather than arithmetic. Engines recover decode efficiency by
batching the steps of many requests and rebuilding the batch at every
iteration~\cite{yu2022orca}.

Both phases rely on the KV cache, which holds the attention keys and values
of every token processed so far. Its size is easy to underestimate.
Llama-3.1-8B has 32 layers and 8 key-value heads of dimension 128; at 16 bits
per value each token occupies 128~KiB, so a single 4,096-token prompt holds
512~MiB. On a 24~GB NVIDIA L4 the weights take roughly 16~GB and leave at most
about 8~GB for cached state. Memory, not arithmetic, bounds how many requests
one GPU can serve at once.

\subsection{Reusing a shared prefix}
\label{sec:bg-prefix}

\begin{figure}[t]
  \centering
  \includegraphics{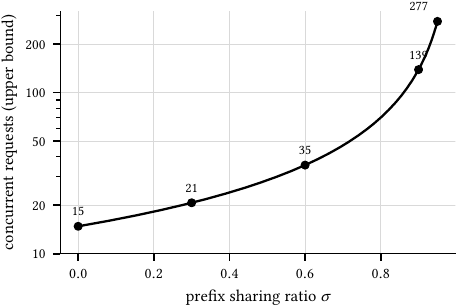}
  \caption{Upper bound on concurrent 4,096-token requests whose KV state fits
    in the memory an NVIDIA L4 has left after the weights of Llama-3.1-8B, as
    a function of the prefix sharing ratio $\sigma$. The shared prefix is
    stored once. Each request adds only its suffix. Computed from the model's
    dimensions rather than measured; activations and engine overhead lower the
    real budget.}
  \label{fig:kvbudget}
\end{figure}

Requests that begin with the same tokens produce identical KV entries for
those tokens, so an engine can store them once. Figure~\ref{fig:kvbudget}
shows what that is worth on our hardware. Without sharing, at most 15
requests of 4,096 tokens fit in the memory the L4 has left. If 90\% of every
prompt is a common prefix stored once, 139 fit. The bound ignores activations
and engine overhead, so the real budget is lower, but its shape survives the
correction: reuse gains little at modest sharing and a great deal at the
sharing levels that agent workloads actually exhibit.

The two engines we study find shared prefixes at different granularities.
vLLM's PagedAttention stores the cache in fixed blocks of 16 tokens, and its
automatic prefix caching reuses a block when a hash of its tokens, together
with the tokens before it, matches a cached block~\cite{kwon2023vllm}. Reuse
is recognised in whole blocks. SGLang's RadixAttention keeps cached sequences
in a radix tree and matches each new prompt against the tree token by
token~\cite{zheng2024sglang}. It therefore recovers a shared prefix of any
length, including one that ends inside a block, and it can evict cached
state one token at a time. Granularity is the property most often credited
when SGLang outperforms vLLM on prefix-heavy traffic.

\subsection{Chunked prefill}
\label{sec:bg-chunked}

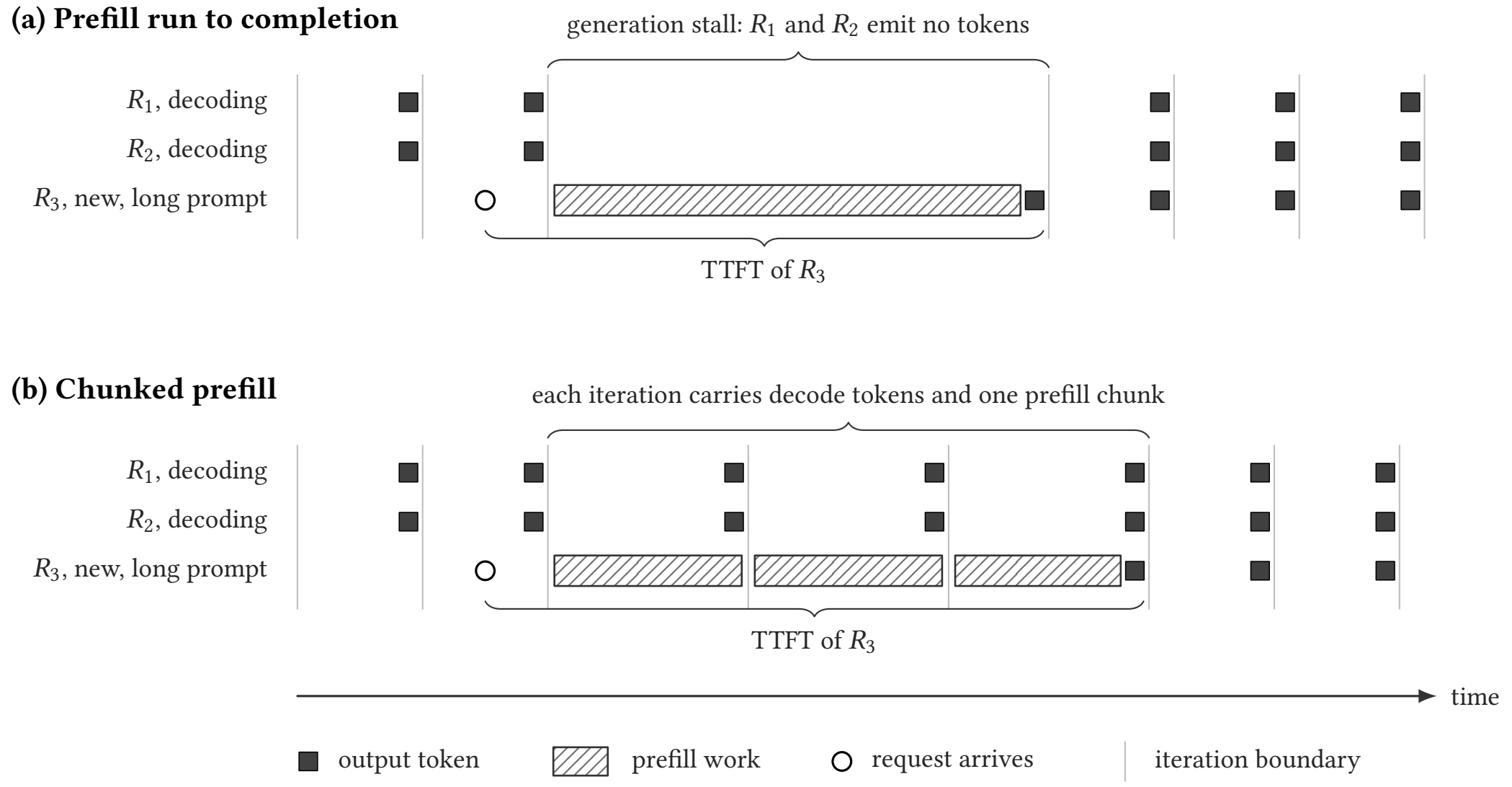
\begin{figure}[t]
  \centering
  \begin{tikzpicture}[paperfig, x=1cm, y=1cm]
  \def\u{1.15}                      %
  \def\ya{1.3}\def\yb{0.85}\def\yc{0.4}   %
  \tikzset{tok/.style={token, rectangle, minimum size=0.17cm, inner sep=0pt},
           arr/.style={circle, draw=black, fill=white, line width=0.6pt,
                       minimum size=0.17cm, inner sep=0pt}}

  \begin{scope}[shift={(2.75,3.4)}]
    \node[title, anchor=west] at (-2.75,2.05) {(a) Prefill run to completion};
    \foreach \t in {0,1,2,6,7,8,9}
      \draw[iterline] (\t*\u,0.05) -- (\t*\u,1.55);
    \foreach \y/\l in {\ya/{$R_1$, decoding}, \yb/{$R_2$, decoding},
                       \yc/{$R_3$, new, long prompt}}
      \node[lbl, anchor=east] at (-0.15,\y) {\l};
    \foreach \t in {1,2,7,8,9} {
      \node[tok] at (\t*\u-0.13,\ya) {};
      \node[tok] at (\t*\u-0.13,\yb) {};
    }
    \path[prefill] (2*\u+0.06,\yc-0.14) rectangle (6*\u-0.26,\yc+0.14);
    \foreach \t in {6,7,8,9} \node[tok] at (\t*\u-0.13,\yc) {};
    \node[arr] at (1.5*\u,\yc) {};
    \draw[decorate, decoration={brace, amplitude=4pt}, draw=black!80]
      (2*\u,1.62) -- (6*\u,1.62)
      node[midway, above=4pt, lbl] {generation stall: $R_1$ and $R_2$ emit no tokens};
    \draw[decorate, decoration={brace, mirror, amplitude=4pt}, draw=black!80]
      (1.5*\u,0.12) -- (6*\u-0.05,0.12)
      node[midway, below=4pt, lbl] {TTFT of $R_3$};
  \end{scope}

  \begin{scope}[shift={(2.75,0)}]
    \node[title, anchor=west] at (-2.75,2.05) {(b) Chunked prefill};
    \foreach \t in {0,1,2,3.6,5.2,6.8,7.8,8.8}
      \draw[iterline] (\t*\u,0.05) -- (\t*\u,1.55);
    \foreach \y/\l in {\ya/{$R_1$, decoding}, \yb/{$R_2$, decoding},
                       \yc/{$R_3$, new, long prompt}}
      \node[lbl, anchor=east] at (-0.15,\y) {\l};
    \foreach \t in {1,2,3.6,5.2,6.8,7.8,8.8} {
      \node[tok] at (\t*\u-0.13,\ya) {};
      \node[tok] at (\t*\u-0.13,\yb) {};
    }
    \path[prefill] (2*\u+0.06,\yc-0.14)   rectangle (3.6*\u-0.06,\yc+0.14);
    \path[prefill] (3.6*\u+0.06,\yc-0.14) rectangle (5.2*\u-0.06,\yc+0.14);
    \path[prefill] (5.2*\u+0.06,\yc-0.14) rectangle (6.8*\u-0.26,\yc+0.14);
    \foreach \t in {6.8,7.8,8.8} \node[tok] at (\t*\u-0.13,\yc) {};
    \node[arr] at (1.5*\u,\yc) {};
    \draw[decorate, decoration={brace, amplitude=4pt}, draw=black!80]
      (2*\u,1.62) -- (6.8*\u,1.62)
      node[midway, above=4pt, lbl] {each iteration carries decode tokens and one prefill chunk};
    \draw[decorate, decoration={brace, mirror, amplitude=4pt}, draw=black!80]
      (1.5*\u,0.12) -- (6.8*\u-0.05,0.12)
      node[midway, below=4pt, lbl] {TTFT of $R_3$};
    \draw[plain] (0,-0.75) -- (9.1*\u,-0.75) node[lbl, right] {time};
  \end{scope}

  \begin{scope}[shift={(2.75,-1.35)}]
    \node[tok] at (0.1,0) {};
    \node[lbl, anchor=west] at (0.25,0) {output token};
    \path[prefill] (2.35,-0.13) rectangle (2.85,0.13);
    \node[lbl, anchor=west] at (2.95,0) {prefill work};
    \node[arr] at (5.0,0) {};
    \node[lbl, anchor=west] at (5.15,0) {request arrives};
    \draw[iterline] (7.6,-0.18) -- (7.6,0.18);
    \node[lbl, anchor=west] at (7.75,0) {iteration boundary};
  \end{scope}
\end{tikzpicture}
  \caption{Prefill scheduling, illustrative and not to scale. (a)~A long
    prefill runs as one iteration, and the running requests $R_1$ and $R_2$
    emit no tokens until it ends. (b)~The same prefill is split into chunks,
    each sharing an iteration with decode work. The running requests keep
    emitting tokens at a slower rhythm, and the first token of $R_3$ arrives
    later than in~(a).}
  \label{fig:timeline}
\end{figure}

A long prefill creates a scheduling problem that has nothing to do with the
cache. If an engine runs a prefill to completion in one iteration, every
request already decoding waits until it ends. Agrawal et al.\ call this a
\emph{generation stall}~\cite{agrawal2024sarathi}. It grows with prompt
length. Their remedy, \emph{chunked
prefill}, splits a prefill of $L$ tokens into $\lceil L/c \rceil$ chunks under
a per-iteration token budget $c$. Each iteration first admits the decode
tokens of running requests and then fills the remaining budget with one chunk
of a waiting prefill (Figure~\ref{fig:timeline}).

The trade is specific, and it is easy to state it wrongly. Running requests
no longer stall, but each of their decode steps now shares an iteration with
prefill work and takes longer. The chunked request gains nothing. Its prefill
spans several iterations, so its first token arrives somewhat later than it
would after one uninterrupted step. Chunking pays off under load: mixed
iterations keep the GPU busy, which raises the throughput an engine can
sustain and shortens the queue that each new request joins. Section~\ref{sec:results}
measures exactly this effect.

We did not build this mechanism. Both engines adopted it. SGLang implements it
in its scheduler and vLLM exposes it as a command-line option,
\texttt{-{}-enable-chunked-prefill}, each following Agrawal et al. The engines differ in the
default. vLLM 0.6.3 enables chunked prefill only when the configured context
length exceeds 32,768 tokens. SGLang 0.4.1 enables it, with a 2,048-token
chunk, on any GPU with less than 25~GB of memory. At our 16,384-token context
on a 24~GB GPU, SGLang chunks and vLLM does not (Section~\ref{sec:system}).

\subsection{Request ordering}
\label{sec:bg-order}

A separate choice decides which waiting request is admitted when capacity
frees. vLLM admits requests in arrival order. SGLang's default policy,
longest prefix match, first admits the waiting request with the most tokens
already cached, which favours the requests whose prefill is cheapest. This
choice is independent of both chunking and cache granularity. It returns in
Section~\ref{sec:results} as the leading candidate for the difference that
survives once chunking is matched.

\section{System Model and Testbed}
\label{sec:system}

\subsection{Requirements}
\label{sec:sys-req}

A difference between two engines can be attributed to a cause only if
everything else is held fixed. We designed the apparatus around three
requirements.
\begin{enumerate}[label=\textbf{Req.~\arabic*},leftmargin=3.6em,itemsep=2pt,topsep=4pt]
  \item \textbf{Identical conditions.} Every run sees the same hardware, the
        same model and, for a given seed, the same request stream.
  \item \textbf{Known configuration.} The configuration of each engine is
        taken from what the engine resolved at start-up, not from the flags we
        passed to it.
  \item \textbf{Reproducibility.} Every request stream is derived from a seed,
        and every measurement can be repeated on a freshly provisioned
        testbed.
\end{enumerate}
The second requirement proved decisive. Both engines resolve their
chunked-prefill setting from the hardware and the configured context length,
so the requested and the effective configuration can differ.

\subsection{Testbed architecture}
\label{sec:sys-arch}

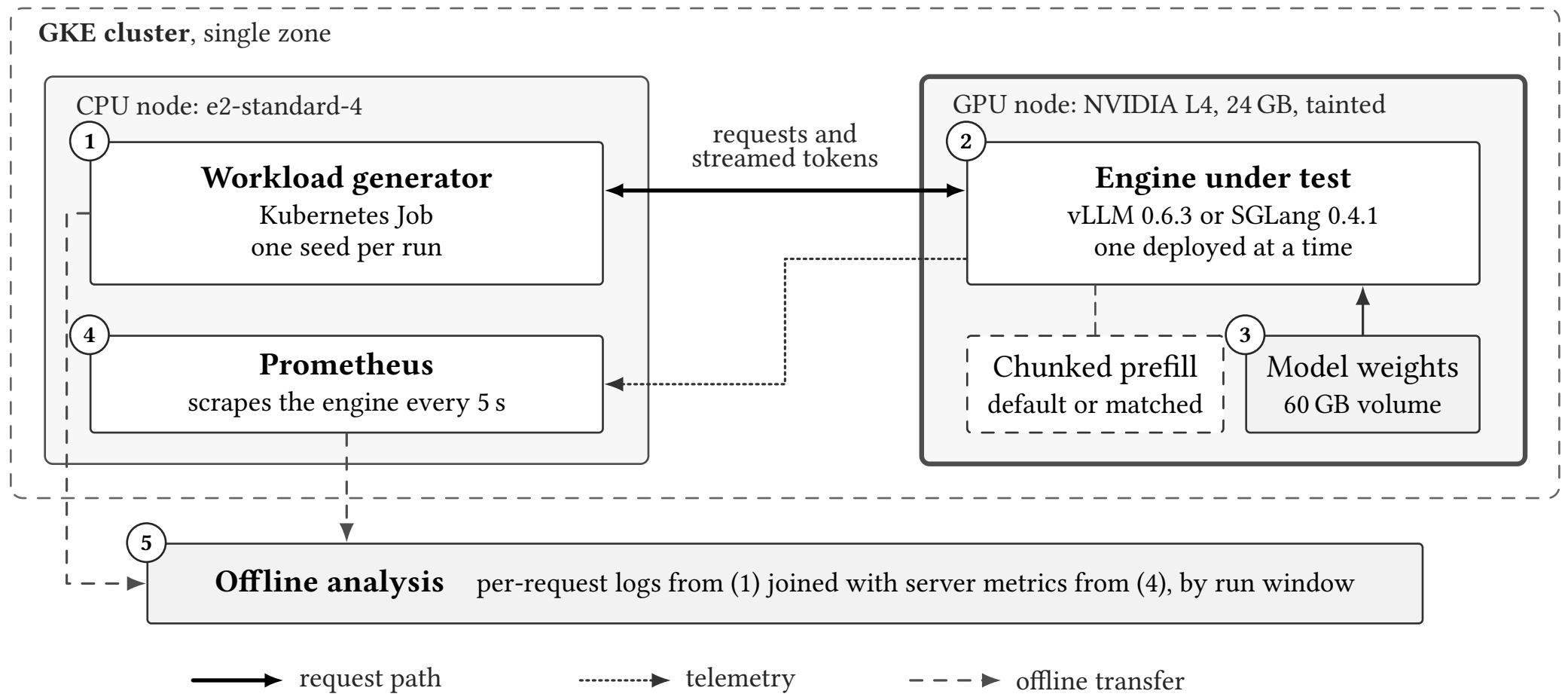
\begin{figure}[t]
  \centering
  \begin{tikzpicture}[paperfig, x=1cm, y=1cm]
  \draw[cluster] (0,1.45) rectangle (13.6,5.75);
  \node[lbl, anchor=west] at (0.12,5.52) {\textbf{GKE cluster}, single zone};
  \draw[region] (0.3,1.75) rectangle (5.6,5.15);
  \node[lbl, anchor=west] at (0.45,4.9) {CPU node: e2-standard-4};
  \draw[regionstrong] (8.0,1.75) rectangle (13.3,5.15);
  \node[lbl, anchor=west] at (8.15,4.9) {GPU node: NVIDIA L4, 24\,GB, tainted};

  \node[comp, minimum width=4.5cm, minimum height=1.25cm] (gen) at (2.95,3.95)
    {\textbf{Workload generator}\\[-1pt]{\scriptsize Kubernetes Job}\\[-2pt]{\scriptsize one seed per run}};
  \node[comp, minimum width=4.5cm, minimum height=0.85cm] (prom) at (2.95,2.45)
    {\textbf{Prometheus}\\[-1pt]{\scriptsize scrapes the engine every 5\,s}};
  \node[comp, minimum width=4.5cm, minimum height=1.25cm] (eng) at (10.65,3.95)
    {\textbf{Engine under test}\\[-1pt]{\scriptsize vLLM 0.6.3 or SGLang 0.4.1}\\[-2pt]{\scriptsize one deployed at a time}};
  \node[setting, minimum width=2.25cm, minimum height=0.85cm] (set) at (9.525,2.45)
    {Chunked prefill\\[-1pt]{\scriptsize default or matched}};
  \node[compshade, minimum width=2.05cm, minimum height=0.85cm] (vol) at (11.875,2.45)
    {Model weights\\[-1pt]{\scriptsize 60\,GB volume}};
  \node[compshade, minimum width=11.2cm, minimum height=0.7cm] (col) at (6.8,0.7)
    {\textbf{Offline analysis}\quad{\scriptsize per-request logs from (1) joined with server metrics from (4), by run window}};

  \draw[reqpath, <->] (gen.east |- 0,4.15) -- (eng.west |- 0,4.15);
  \node[lbl, align=center] at (6.8,4.55) {requests and\\[-2pt]streamed tokens};
  \draw[telemetry] (eng.west |- 0,3.55) -- (6.8,3.55) -- (6.8,2.45) -- (prom.east);
  \draw[densely dashed, draw=black!70, line width=0.5pt] (set.north) -- (set.north |- eng.south);
  \draw[plain] (vol.north) -- (vol.north |- eng.south);
  \draw[offline] (prom.south) -- (prom.south |- col.north);
  \draw[offline] (gen.west) -- ++(-0.2,0) |- (col.west);

  \foreach \n/\c in {1/gen, 2/eng, 3/vol, 4/prom, 5/col}
    \node[num] at (\c.north west) {\n};

  \draw[reqpath]   (1.6,-0.15) -- (2.4,-0.15) node[lbl, right] {request path};
  \draw[telemetry] (5.0,-0.15) -- (5.8,-0.15) node[lbl, right] {telemetry};
  \draw[offline]   (7.9,-0.15) -- (8.7,-0.15) node[lbl, right] {offline transfer};
\end{tikzpicture}
  \caption{Testbed architecture. The workload generator~(1) drives the engine
    under test~(2), which loads its weights from a persistent volume~(3).
    Prometheus~(4) scrapes the engine every 5\,s. After each run, an offline
    step~(5) joins the generator's per-request records with the engine's
    metrics by run window. Line style carries meaning: thick solid lines are
    the request path, dotted lines telemetry, dashed lines offline transfer.
    The thick border marks the tainted GPU node.}
  \label{fig:testbed}
\end{figure}

We ran every experiment on Google Kubernetes Engine (Kubernetes 1.35) in a
single zone of the \texttt{us-central1} region. Terraform provisions the whole
environment in two stages: the first creates the cluster, its node pools and a
container registry, and the second installs the platform. Because the
environment exists only as code, we destroyed it between campaigns and rebuilt
it identically. Figure~\ref{fig:testbed} shows the components; the numbers
below refer to it.

The cluster has two node pools with separate roles. The GPU pool holds a
single \texttt{g2-standard-8} instance with one NVIDIA L4 GPU (24~GB) and GKE's
managed driver. A \texttt{NoSchedule} taint keeps every workload except the
engine under test~(2) off this node. We held the pool at one node for the
duration of each campaign. Releasing and reacquiring the GPU between runs
would have exposed the campaign to regional capacity shortages and left gaps
in paired measurements. In the matched campaign the node was a Spot VM, which
the provider may reclaim at any time. It did so twice. Each time the pool
replaced the node within 20 minutes, and Section~\ref{sec:des-protocol}
describes how we treated the conditions that the reclamation interrupted.

The CPU pool holds an always-on \texttt{e2-standard-4} node for the monitoring
stack, the cluster's control components and the workload generator~(1).
Placing the generator there removes two sources of error. It cannot compete
with the engine's own CPU-side scheduler, and its requests cross only the
cluster network, not a public round trip.

The second Terraform stage installs Prometheus~(4) with a 5\,s scrape
interval, cert-manager, KServe in \texttt{RawDeployment} mode, and KEDA. KServe
exposes the engine as a plain Kubernetes Deployment and Service. KEDA is
installed but plays no part in this study. We disabled autoscaling for every
measurement, and the engine ran as exactly one replica throughout each sweep.
Model weights reside on a 60~GB persistent volume~(3), so an engine restart
reloads them from local disk rather than from a remote registry. After each
run, an offline step~(5) retrieves the engine's metrics for that run's time
window and joins them with the generator's per-request records.

\subsection{Systems under test}
\label{sec:sys-engines}

We compare vLLM~0.6.3~\cite{kwon2023vllm} and SGLang~0.4.1~\cite{zheng2024sglang}.
Both run without modification from their published container images,
\texttt{vllm/vllm-openai:v0.6.3} and \texttt{lmsysorg/sglang:v0.4.1-cu121},
and both serve Llama-3.1-8B-Instruct in \texttt{bfloat16} with a 16,384-token
context and prefix caching enabled. Image digests, which
unlike tags cannot be reassigned, are listed in Appendix~\ref{app:artifact}.

We chose the model for its KV-cache footprint. With 32 layers and eight
key-value heads of dimension 128, it stores 128~KiB per token. A 7B model
with four key-value heads and 28 layers, such as Qwen2.5-7B, stores 56~KiB.
The larger footprint makes the cache fill, and eviction pressure become
reachable, within the load a single 24~GB GPU can sustain.

Table~\ref{tab:config} lists the resolved configuration of each engine, with
every value read from the engine's own start-up log rather than from the
flags we passed. vLLM runs its V0 engine
with automatic prefix caching over 16-token blocks. SGLang runs RadixAttention
with the FlashInfer attention backend, its overlap scheduler, and
longest-prefix-match ordering of waiting requests.

\begin{table}[t]
  \centering
  \caption{Resolved engine configurations, read from the deployed manifests
    and the engines' start-up logs. Only vLLM's chunked-prefill setting
    differs between the two arms. The
    rows above the rule are identical in both arms but differ between the
    engines; they bound what the gap remaining under matched configuration can
    be attributed to (Section~\ref{sec:threats}).}
  \label{tab:config}
  \small
  \begin{tabular}{lll}
    \toprule
                             & vLLM 0.6.3                     & SGLang 0.4.1 \\
    \midrule
    Prefix cache             & PagedAttention with APC        & RadixAttention \\
    Match unit               & 16-token block                 & token \\
    Memory budget            & \texttt{gpu\_memory\_utilization} 0.88 & \texttt{mem\_fraction\_static} 0.85 \\
    Attention backend        & engine default                 & FlashInfer \\
    Order of waiting requests & arrival order                 & longest prefix match \\
    \midrule
    \multicolumn{3}{l}{\emph{Chunked prefill}} \\
    \quad vendor defaults    & off                            & on, 2,048 tokens \\
    \quad matched            & on, 2,048 tokens               & on, 2,048 tokens \\
    \bottomrule
  \end{tabular}
\end{table}

\subsection{Configurations}
\label{sec:sys-configs}

Each engine runs in two configurations.

\paragraph{Vendor defaults.} Each engine runs with the chunked-prefill
behaviour its authors ship. vLLM enables chunked prefill only when the
configured context length exceeds 32,768 tokens, so at 16,384 tokens it runs
each prefill as a single step. Our manifest states this value explicitly with
\texttt{-{}-enable-chunked-prefill=False}, which pins vLLM's own resolution
rather than changing it. SGLang finds less than 25~GB of GPU memory and
resolves a 2,048-token chunk; we pass it no chunking flag. This is the
deployment a practitioner following each project's documentation obtains.

\paragraph{Matched chunked prefill.} We enable chunked prefill on vLLM and set
its per-iteration token budget to 2,048 tokens, passing
\texttt{-{}-enable-chunked-prefill} and \texttt{-{}-max-num-batched-tokens=2048}.
vLLM's start-up log confirms that the setting took effect: it reports chunked
prefill enabled with a 2,048-token budget. SGLang needs no change. Its rate
runs used the same manifest as in the defaults arm, and for the 8,192-token
runs we passed \texttt{-{}-chunked-prefill-size=2048} explicitly and recorded
the resolved value from that pod's start-up log. Both resolve to the same
2,048-token chunk. The manipulation therefore changes one setting of one
engine, and SGLang serves as an internal control.

Matching the chunk size does not make the engines identical. The rows above
the rule in Table~\ref{tab:config} remain different, and one of them needs
care in interpretation. The two memory-budget parameters measure different
things. vLLM's bounds the fraction of GPU memory the whole engine may use;
SGLang's sets the fraction reserved for weights and static state. Equal-looking
values therefore need not give equal KV-cache capacity.

\subsection{Workload model}
\label{sec:sys-workload}

\begin{figure}[t]
  \centering
  \begin{tikzpicture}[paperfig, x=1cm, y=1cm,
    dagnode/.style={circle, draw=black!80, fill=white, line width=0.5pt,
                    minimum size=0.38cm, inner sep=0pt, font=\tiny}]
  \draw[cluster] (0,2.05) rectangle (13.85,5.5);
  \node[lbl, anchor=west] at (0.12,5.27)
    {\textbf{Offline compiler}: one seed fixes arrivals, DAGs and prompts};

  \begin{scope}[shift={(0.15,0)}]
    \draw[comp] (0,2.2) rectangle (4.3,5.0);
    \node[title, anchor=west] at (0.15,4.72) {L1\ \ Arrivals};
    \node[lbl, anchor=west] at (0.15,4.38) {Gamma process, CV 1.5};
    \draw[plain] (0.3,2.85) -- (4.0,2.85);
    \node[lbl, anchor=north east] at (4.0,2.78) {time};
    \foreach \t in {0.5,0.62,0.71,0.8,1.65,2.6,2.69,2.77,2.85,3.75}
      \draw[line width=0.9pt] (\t,2.85) -- (\t,3.35);
    \node[note] at (2.15,3.75) {bursts and quiet gaps};
  \end{scope}

  \begin{scope}[shift={(4.7,0)}]
    \draw[comp] (0,2.2) rectangle (4.3,5.0);
    \node[title, anchor=west] at (0.15,4.72) {L2\ \ Agent DAG};
    \node[lbl, anchor=west] at (0.15,4.38) {planner, 4 workers, aggregator};
    \node[dagnode] (p) at (0.55,3.18) {P};
    \foreach \i/\y in {1/3.85, 2/3.4, 3/2.95, 4/2.5}
      \node[dagnode] (w\i) at (2.15,\y) {W};
    \node[dagnode] (a) at (3.75,3.18) {A};
    \foreach \i in {1,...,4} {
      \draw[plain] (p) -- (w\i);
      \draw[plain] (w\i) -- (a);
      \path (p) -- (w\i) coordinate[pos=0.55] (m\i);
      \fill[gray!45, draw=black!70, line width=0.3pt]
        ($(m\i)+(-0.13,-0.045)$) rectangle ($(m\i)+(0.13,0.045)$);
    }
  \end{scope}

  \begin{scope}[shift={(9.25,0)}]
    \draw[comp] (0,2.2) rectangle (4.3,5.0);
    \node[title, anchor=west] at (0.15,4.72) {L3\ \ Prompt};
    \node[lbl, anchor=west] at (0.15,4.38) {shared prefix, unique suffix};
    \draw[draw=black!80, fill=gray!30, line width=0.5pt] (0.3,3.0) rectangle (3.05,3.55);
    \node[lbl] at (1.675,3.275) {prefix, $F$ tokens};
    \draw[draw=black!80, fill=white, line width=0.5pt] (3.05,3.0) rectangle (4.0,3.55);
    \node[lbl] at (3.525,3.275) {$D$};
    \node[lbl] at (2.15,2.6) {$\sigma = F/(F+D)$};
    \node[note] at (2.15,3.85) {prefix identical across requests};
  \end{scope}

  \node[compshade, minimum width=2.6cm, minimum height=0.85cm] (tr) at (1.45,1.0)
    {Compiled trace\\[-1pt]{\scriptsize arrivals, DAGs, prompts}};
  \node[comp, minimum width=2.6cm, minimum height=0.85cm] (val) at (4.95,1.0)
    {Validator\\[-1pt]{\scriptsize six checks}};
  \node[comp, minimum width=2.6cm, minimum height=0.85cm] (dis) at (8.45,1.0)
    {Dispatcher\\[-1pt]{\scriptsize asynchronous, streaming}};
  \node[comp, minimum width=2.6cm, minimum height=0.85cm] (eng) at (11.95,1.0)
    {Engine under test};
  \draw[plain] (1.45,2.05) -- (tr.north);
  \draw[plain] (tr) -- (val);
  \draw[plain] (val) -- node[lbl, above] {pass} (dis);
  \draw[reqpath] (dis) -- (eng);
  \draw[plain, dashed] (val.south) -- (4.95,0.12);
  \node[note, anchor=north] at (4.95,0.1) {fails a check: trace discarded};
\end{tikzpicture}
  \caption{Workload generator. An offline compiler produces a complete trace
    from one seed: an arrival process~(L1) schedules agent DAGs, a topology
    layer~(L2) expands each into dependent requests, and a prompt layer~(L3)
    builds each request from a shared prefix and a unique suffix. Gray bars in
    L2 mark the simulated tool call that precedes each worker. A trace that
    fails validation is discarded before any request is sent.}
  \label{fig:generator}
\end{figure}
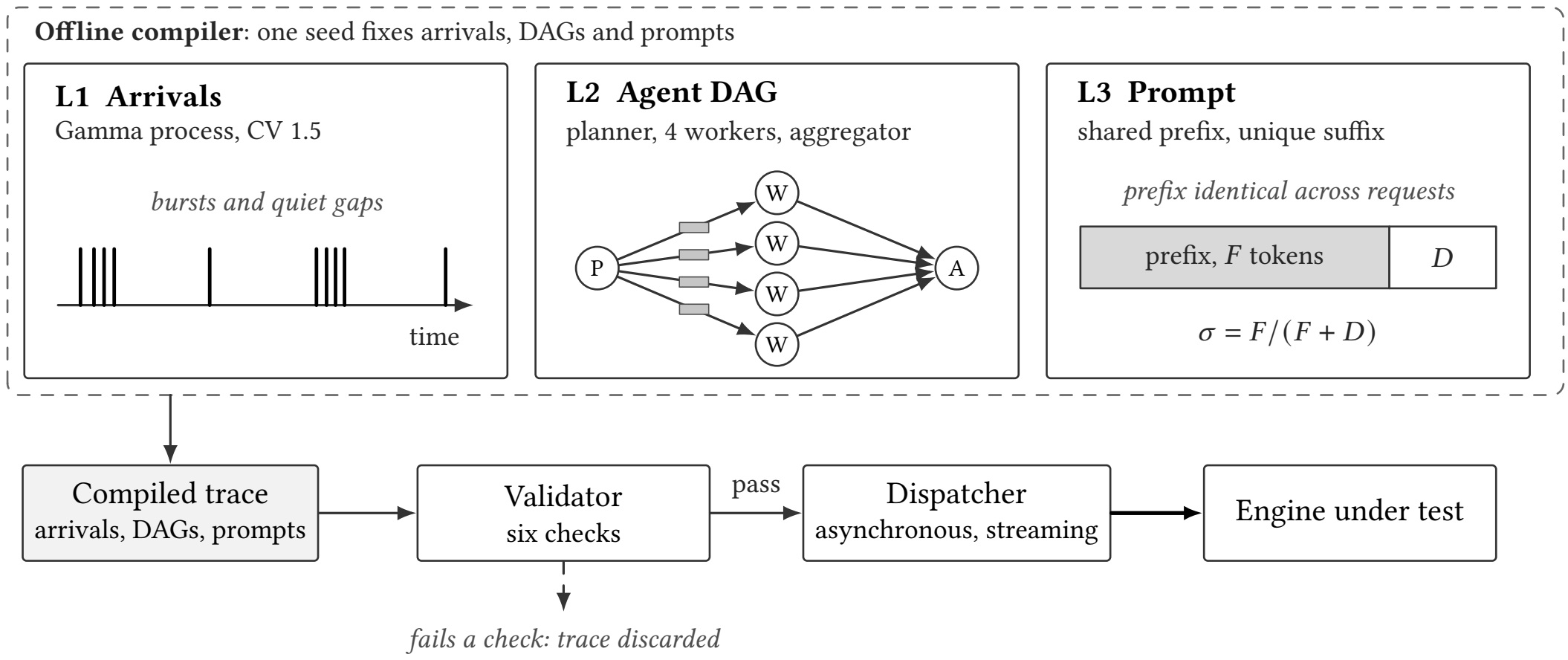

Multi-agent applications issue requests as dependency graphs whose members
share a long prompt prefix and differ in a short suffix. Our generator
reproduces that structure under exact control. It compiles a complete trace
offline, so that at run time the dispatcher only waits, sends and records
(Figure~\ref{fig:generator}). The trace has three layers.

\paragraph{Arrivals (L1).} Agent graphs arrive at times $t_1 < t_2 < \dots$
whose gaps are independent draws from a Gamma distribution with mean
$1/\lambda$ and coefficient of variation $c_v = 1.5$, that is, with shape
$k = 1/c_v^2 \approx 0.444$ and scale $\theta = c_v^2/\lambda$. A coefficient of
variation above one produces bursts separated by quiet gaps, which matches
the burstiness reported for production LLM traffic~\cite{wang2024burstgpt,xiang2025servegen}. We
express $\lambda$ in agent graphs per second.

\paragraph{Topology (L2).} Each arrival instantiates a graph $G = (V, E)$ with
$V = \{p, w_1, \dots, w_4, a\}$ and
$E = \{(p, w_i)\} \cup \{(w_i, a)\}$ for $i = 1, \dots, 4$: a planner $p$, four
workers, and an aggregator $a$. A request is issued once all of its
predecessors have completed. Each worker $w_i$ additionally waits for a
simulated tool call whose duration $\tau_i$ is drawn from published
Model Context Protocol tool latencies~\cite{bian2025tokencake}, so the
workers do not start in lock-step. Every graph contains six requests, and the
request rate is $6\lambda$. Output length is capped at 128 tokens for the
planner and the workers and at 256 tokens for the aggregator.

\paragraph{Prompts (L3).} Every prompt is a fixed prefix of $F$ tokens,
identical across all requests of a run, followed by a suffix of $D$ tokens
unique to the request. The prompt length is $L = F + D$, and the prefix
sharing ratio
\begin{equation}
  \sigma = \frac{F}{F + D}
  \label{eq:sigma}
\end{equation}
is our principal independent variable. We sample tokens uniformly from the
model's vocabulary, excluding special tokens. Decoding a token sequence to
text and encoding it again need not return the same sequence, so the
generator corrects each segment iteratively until the model's own tokenizer
returns exactly the requested count.

Synthetic tokens are deliberate. Prefix-cache matching compares token
identifiers, not meaning, so two byte-identical sequences share cache entries
whatever they say. Sampled tokens give exact control over $\sigma$ and $L$,
which no natural corpus provides. What they omit, realistic length
distributions and near-duplicate prompts, is discussed in
Section~\ref{sec:threats}.

\paragraph{Validation and dispatch.} Before any request is sent, a validator
checks six properties of the compiled trace: the prefix's SHA-256 digest
matches the trace manifest, the achieved $\sigma$ lies within 1\% of its
target, every suffix is distinct, every graph is acyclic and resolvable,
arrival times are ordered, and tool delays are non-negative and bounded. A
trace that fails any check is discarded. The check matters because a subtly
broken prefix still produces latency numbers; they simply measure an empty
cache. A dispatcher then replays the trace asynchronously, sending each
request as a streaming completion and timestamping every returned chunk.
Section~\ref{sec:metrics} defines the metrics computed from those timestamps.

\section{Performance Metrics}
\label{sec:metrics}

We define metrics at three levels: individual requests, individual runs, and
comparisons between the two engines. All are computed from the dispatcher's
per-chunk timestamps and the run's server-side telemetry, with no
instrumentation inside either engine.

\subsection{Request metrics}
\label{sec:met-request}

For request $r$, let $s_r$ be the time the dispatcher sends it and
$c_{r,0} < c_{r,1} < \dots < c_{r,n_r-1}$ the arrival times of its $n_r$
output tokens. Both engines stream one token per chunk in our configuration,
so chunk and token timestamps coincide. The \emph{time to first token} is
\begin{equation}
  \mathrm{TTFT}_r = c_{r,0} - s_r .
  \label{eq:ttft}
\end{equation}
It includes the time the request waits for admission and the time its
prefill takes. A request that queues behind other work therefore shows a
long TTFT even when the engine processes it quickly once admitted, and this
property drives several results in Section~\ref{sec:results}. The \emph{time
per output token} averages the gaps between the remaining tokens:
\begin{equation}
  \mathrm{TPOT}_r = \frac{c_{r,n_r-1} - c_{r,0}}{n_r - 1}, \qquad n_r \ge 2 .
  \label{eq:tpot}
\end{equation}
TPOT measures decode speed. Agrawal et al.\ report the same per-step
quantity as time between tokens~\cite{agrawal2024sarathi}; we use TPOT, the
name both engines use in their own metrics.

\subsection{Run metrics}
\label{sec:met-run}

A run serves one engine, in one configuration, one trace. Its measured
requests exclude the first agent graph, which pays connection costs that
later graphs do not (Section~\ref{sec:design}). For each run we report the
median (p50) and 95th percentile (p95) of TTFT and the median of TPOT over
its measured requests.

The two percentiles answer different questions. A median that moves means
most requests changed; a 95th percentile that moves while the median does not
means a minority of requests waited much longer than the rest. Section~\ref{sec:results}
relies on that distinction. We do not report the 99th percentile. Our
lightest runs contain under two hundred measured requests, so a 99th
percentile would rest on one or two observations.

Let $N_{\mathrm{ok}}$ be the number of requests a run completes and $W$ its
wall-clock duration. The \emph{achieved throughput} is
\begin{equation}
  X = N_{\mathrm{ok}} / W ,
  \label{eq:throughput}
\end{equation}
and the \emph{offered load} is $\Lambda = 6\lambda$ requests per second, since
every agent graph contains six requests (Section~\ref{sec:sys-workload}). An
engine \emph{keeps pace} with the offered load when $X \ge 0.97\,\Lambda$.
The tolerance absorbs the variability of bursty arrivals. A run of finite
length realises the nominal rate only approximately. At the lightest load a
run contains about 45 agent graphs, and a single run's achieved throughput
exceeds $\Lambda$ by up to 29\%. The deviation affects both engines alike,
since both receive the identical trace. An engine that falls
short of the threshold accumulates a queue over the run, and its TTFT then
grows with the run's length rather than reflecting the engine's speed.
\emph{Capacity} is the largest offered load at which an engine keeps pace.

From the engine's telemetry we take the peak fraction of KV-cache memory in
use during the run, which we call \emph{KV-cache utilisation}, and the number
of requests the engine preempted.

\subsection{Comparing engines}
\label{sec:met-compare}

The two engines receive identical request streams for a given seed
(Section~\ref{sec:sys-workload}). A \emph{condition} $c$ fixes the arrival
rate, the sharing ratio and the prompt length; a \emph{repetition} $k$ fixes
the seed. For a latency metric $M$, the engine gap in repetition $k$ is the
paired ratio
\begin{equation}
  R_{c,k} = \frac{M_{c,k}^{\mathrm{vLLM}}}{M_{c,k}^{\mathrm{SGLang}}} ,
  \label{eq:ratio}
\end{equation}
so $R > 1$ means SGLang is faster. For throughput we invert the ratio,
$X^{\mathrm{SGLang}} / X^{\mathrm{vLLM}}$, so that $R > 1$ again favours SGLang.

Ratios are multiplicative quantities, and we aggregate them as such. Over
$n$ repetitions, the reported gap is the geometric mean
\begin{equation}
  \bar{R}_c = \exp\!\Big(\frac{1}{n}\sum_{k=1}^{n} \ln R_{c,k}\Big) ,
  \label{eq:gmean}
\end{equation}
following Fleming and Wallace~\cite{fleming1986lie}. The arithmetic mean of
ratios is biased upward and depends on which engine is placed in the
numerator. The geometric mean has neither defect, and because the runs are
paired it equals the ratio of the two engines' geometric means exactly. Every
absolute value we report is therefore consistent with every ratio. We
aggregate absolute latencies and throughputs across repetitions the same
way. KV-cache utilisation is a bounded proportion, not a multiplicative
quantity, so for it we use the arithmetic mean.

\subsection{Decomposing the gap}
\label{sec:met-decomp}

Let $\bar{R}_c^{\mathrm{def}}$ and $\bar{R}_c^{\mathrm{mat}}$ be the gap in
condition $c$ under vendor defaults and under matched chunked prefill. On a
log scale the default gap splits into a part that matching removes and a part
that remains. The \emph{removed fraction}
\begin{equation}
  \phi_c = \frac{\ln\big(\bar{R}_c^{\mathrm{def}} / \bar{R}_c^{\mathrm{mat}}\big)}
                {\ln \bar{R}_c^{\mathrm{def}}}
  \label{eq:removed}
\end{equation}
is the share of the default gap attributable to the chunked-prefill
configuration. At $\phi_c = 1$ matching closes the gap entirely, and at
$\phi_c = 0$ it changes nothing. We compute $\phi_c$ only when the default gap
exceeds 10\% in either direction, $|\ln \bar{R}_c^{\mathrm{def}}| > \ln 1.1$,
because below that threshold the denominator approaches zero and the fraction
swings with noise in either arm. The guard excludes one condition: the median
at 0.12 agent graphs per second, where the default gap is $1.09\times$.

\subsection{Statistical inference}
\label{sec:met-stats}

Every condition has $n = 3$ repetitions. We give 95\% confidence intervals
for $\bar{R}_c$ on the log scale, as
$\exp\!\big(\overline{\ln R} \pm t_{0.975,\,n-1}\, s/\sqrt{n}\big)$, where $s$
is the sample standard deviation of $\ln R_{c,k}$ and $t_{0.975,2} = 4.30$.
These intervals are asymmetric and strictly positive. An arithmetic interval
on a ratio can extend below zero, which has no meaning for a ratio.

We test whether a gap differs from parity with a paired $t$-test on the log
latencies, $d_k = \ln M_{c,k}^{\mathrm{vLLM}} - \ln M_{c,k}^{\mathrm{SGLang}}$,
and report a difference as significant at $p < 0.05$. Pairing removes the
variance that the arrival process contributes to both engines alike. The
effect is large. On the sharing sweep an unpaired test finds a significant
difference at one sharing level of five, while the paired test finds one at
four. Three repetitions give low statistical power, so we treat
a non-significant result as an absence of evidence, never as evidence of
equality.

\section{Experimental Design}
\label{sec:design}

\subsection{Experiment grid}
\label{sec:des-grid}

We ran three sweeps under vendor defaults and a subset of their conditions
again under matched chunked prefill (Table~\ref{tab:sweeps}). Each sweep
varies one factor and holds the others fixed. The \emph{rate sweep} varies
the arrival rate at $\sigma = 0.9$ with 4,096-token prompts. The
\emph{sharing sweep} varies $\sigma$ at a fixed low rate. The
\emph{prompt-length sweep} varies the prompt length at $\sigma = 0.9$.

Every condition ran three times per engine. Each repetition uses its own
seed, and the same seed drives both engines, so the two engines in a
repetition receive byte-identical traces. The matched arm repeats the three
conditions in which the default gap was significant and throughput could be
compared: 0.12 and 0.28 agent graphs per second on the rate sweep, and
8,192-token prompts on the prompt-length sweep. One further matched run, of
vLLM at 0.20 agent graphs per second, has no SGLang counterpart. It appears
as an absolute value in Section~\ref{sec:results} and never in a paired
comparison.

\begin{table}[t]
  \caption{Experiment grid. Every condition ran three times per engine, with
    the seed shared between engines. Rates are in agent graphs per second.}
  \label{tab:sweeps}
  \small
  \begin{tabular}{llll}
    \toprule
    Sweep & Varied & Held fixed & Duration per run \\
    \midrule
    Rate     & $\lambda \in \{0.05, 0.12, 0.20, 0.28, 0.40\}$ & $\sigma = 0.9$, $L = 4{,}096$ & 900\,s if $\lambda \le 0.12$, else 600\,s \\
    Sharing  & $\sigma \in \{0, 0.3, 0.6, 0.9, 0.95\}$ & $\lambda = 0.012$, $L = 4{,}096$ & 2,400\,s \\
    Prompt length & $L \in \{512, 2{,}048, 8{,}192\}$ & $\sigma = 0.9$; $\lambda$ scaled with $L$ & 300, 400, 900\,s \\
    \midrule
    Matched  & rate 0.12 and 0.28; $L = 8{,}192$ & as in the defaults arm & as in the defaults arm \\
    \bottomrule
  \end{tabular}
\end{table}

\subsection{Choosing operating points}
\label{sec:des-points}

A comparison of latencies informs only if both engines run in the regime of
interest, and that regime depends on each engine's capacity. We did not know
the capacities in advance, so we measured them before choosing any rate.

\paragraph{Establishing capacity.} We drove vLLM far past saturation in a
diagnostic run at 1.0 agent graphs per second, with $\sigma = 0.9$ and
4,096-token prompts. Against an offered load of 6.0 requests per second it
completed 1.56, and its queue reached 230 requests. That places vLLM's
capacity near 0.26 agent graphs per second. The same run showed which
resource binds. Peak KV-cache utilisation reached 0.995, so memory rather
than arithmetic limited the engine, and prefix caching could therefore affect
the outcome.

\paragraph{Rate sweep.} We chose five rates that bracket the measured
capacity, from well below the saturation point to well above it. Runs last
900\,s at the two lowest rates and 600\,s above them, so that every run
collects enough requests for stable percentiles.

\paragraph{Sharing sweep.} Capacity depends on $\sigma$. At $\sigma = 0$ every
request prefills all 4,096 tokens, about ten times the work at $\sigma = 0.9$.
A preliminary run at 0.12 agent graphs per second saturated vLLM at $\sigma = 0$
and failed repeatedly. We therefore ran all five sharing levels at a single
rate of 0.012, low enough that even $\sigma = 0$ stays below saturation, and
extended each run to 2,400\,s. Holding the rate fixed across $\sigma$ keeps a
latency difference attributable to prefill work rather than to differing
queue depths.

\paragraph{Prompt-length sweep.} Capacity also falls as prompts grow. We
scaled the rate inversely with length to hold utilisation roughly constant:
1.0, 0.26 and 0.065 agent graphs per second for 512, 2,048 and 8,192 tokens.
A fourth point at 12,288 tokens saturated both engines at its assigned rate
and completed only 78\% of its requests. We excluded it rather than report a
measurement dominated by the queue.

\subsection{Workload realisation}
\label{sec:des-workload}

Table~\ref{tab:workload} reports what the runs contained. The 99 runs
measured 63,288 requests, 48,228 under vendor defaults and 15,060 under
matched chunked prefill. Request counts vary widely within a sweep because
both rate and duration vary; on the sharing sweep, where neither does, they
vary much less. Every run completed at least 99.8\% of its requests.

\begin{table}[t]
  \caption{Workload realisation: measured requests per run, by sweep,
    excluding the first agent graph of every run. Generated from the run logs
    by \texttt{make\_workload\_table.py}.}
  \label{tab:workload}
  \small
  \begin{tabular}{lrrrrrrr}
    \toprule
          &      & \multicolumn{4}{c}{Requests per run} & Requests & Lowest \\
    \cmidrule(lr){3-6}
    Sweep & Runs & Mean & Median & Std.\ dev. & Range & in total & success (\%) \\
    \midrule
    Rate sweep & 30 & 886 & 774 & 418 & 276--1,674 & 26,580 & 99.9 \\
Sharing sweep & 30 & 170 & 144 & 42 & 138--228 & 5,100 & 100.0 \\
Prompt-length sweep & 18 & 919 & 594 & 708 & 240--1,926 & 16,548 & 99.8 \\
Matched arm & 21 & 717 & 720 & 294 & 240--1,104 & 15,060 & 100.0 \\
\midrule
Total & 99 & & & & & 63,288 & 99.8 \\
\bottomrule
  \end{tabular}
\end{table}

The logs confirm the workload's structure as well as its volume. Across all
runs the planner, worker and aggregator requests occur in the ratio 1 : 4 : 1
exactly, as six-request agent graphs require. The dispatcher kept to the
compiled schedule. Its median lag behind a graph's planned start time was
1.1\,ms across runs and at most 12\,ms in any run, negligible against the
latencies we measure.

\subsection{Measurement protocol}
\label{sec:des-protocol}

\paragraph{Campaign control.} We ran each measurement as a Kubernetes Job on
the CPU node. A runner submits the Jobs one at a time, since concurrent runs
would share the GPU. It waits 30\,s between runs so that one run's cache
state does not carry into the next. A run passes only if its log holds a
complete results record and at least 95\% of its requests succeeded. The
runner retries a failing run up to twice, keeps the log of every failed
attempt, and records a run that still fails rather than halting the campaign.
Every run writes its full per-request results to its log, so the runner
resumes after an interruption by skipping completed runs. We deployed one
engine at a time.

\paragraph{Node replacement.} When the provider reclaimed the GPU node during
a run (Section~\ref{sec:sys-arch}), we discarded every repetition of the
interrupted condition, including those already complete, and repeated all
three on the replacement node. Each condition's repetitions therefore share
one node and one cache history. This happened at 0.20 agent graphs per second
for vLLM and at 0.12 for SGLang, both in the matched campaign. The discarded
logs are kept in the artifact and excluded from analysis.

\paragraph{Warm-up.} Before each run the dispatcher polls the engine's health
endpoint and sends 20 requests whose results it discards. Measurement
therefore starts after connections are established and lazily initialised
state has been exercised. CUDA-graph capture happens earlier, during engine
start-up. The first agent graph of every run is also excluded from analysis,
because it pays connection costs that later graphs do not. Requests time out
after 300\,s.

\paragraph{Telemetry.} Prometheus records KV-cache utilisation, queue depth
and preemptions from the engine at 5\,s resolution. Each run logs its start
and end times, and after the run we retrieve the server-side series for
exactly that window and join them to the client-side records by run
identifier.

\paragraph{Restart measurement.} For the restart experiment of
Section~\ref{sec:variability}, a separate script deletes the serving pod, waits
for it to terminate, applies the deployment again, and records the Kubernetes
event timestamps for scheduling, image pull, container start and readiness.
The GPU node stays provisioned throughout, so the measurement isolates engine
initialisation from node provisioning.

\section{Results}
\label{sec:results}

We report geometric means over three seed-paired repetitions, 95\% intervals
on the log scale, and paired $t$-tests on log latencies, all as defined in
Section~\ref{sec:metrics}. Five findings answer the research questions of
Section~\ref{sec:intro}:
\begin{enumerate}[label=\textbf{F\arabic*},leftmargin=2.4em,itemsep=2pt,topsep=4pt]
  \item Under vendor defaults the gap appears only under load and mainly in
        the tail, and SGLang pays for it in decode speed (RQ1,
        Section~\ref{sec:res-defaults}).
  \item One configuration difference, chunked prefill, accounts for 30 to
        67\% of the default latency gap and 78\% of the capacity gap (RQ2,
        Section~\ref{sec:res-decomp}).
  \item Near saturation the latency gap measures queue growth. A fluid queue
        model driven only by the throughput shortfall reproduces it, and
        predicts that it scales with the length of the benchmark (RQ3,
        Section~\ref{sec:res-queue}).
  \item A tail-latency advantage of 2.4 to $6.4\times$ survives matching at
        equal throughput (RQ2, Section~\ref{sec:res-tail}).
  \item The gap shrinks and then reverses as prefix sharing grows, contrary
        to the architectural prediction (RQ4, Section~\ref{sec:res-sigma}).
\end{enumerate}

\subsection{The gap under vendor defaults}
\label{sec:res-defaults}

\begin{figure}[t]
  \centering
  \includegraphics{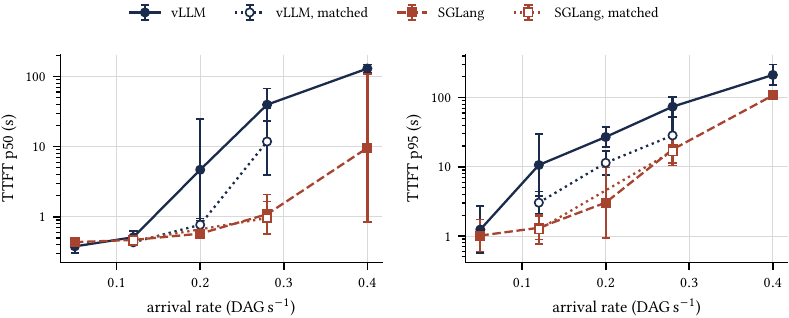}
  \caption{TTFT against arrival rate at $\sigma = 0.9$ with 4,096-token
    prompts. Filled markers with solid or dashed lines show vendor defaults;
    hollow markers with dotted lines show chunked prefill matched at 2,048
    tokens, which we measured at a subset of rates. Bars are 95\% intervals
    on the log scale around geometric means of three runs.}
  \label{fig:rate}
\end{figure}

At the two lowest rates the engines cannot be told apart at the median:
$\bar{R} = 0.87$ and $1.09$, neither significant (Figure~\ref{fig:rate}).
Above them vLLM's median TTFT rises steeply. At 0.28 agent graphs per second
it is $36.4\times$ SGLang's (95\% interval $[13.8, 95.6]$). Read alone, this
suggests that the cache architectures differ by more than an order of
magnitude.

\begin{figure}[t]
  \centering
  \includegraphics{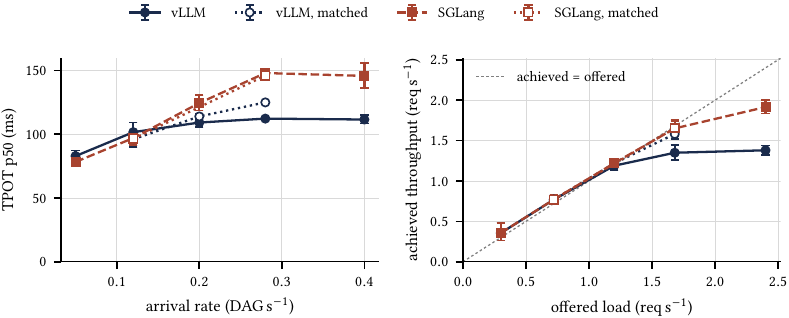}
  \caption{Left: median TPOT against arrival rate. Right: achieved against
    offered load; points below the diagonal mark an engine that does not
    keep pace. Markers and lines as in Figure~\ref{fig:rate}. Geometric means
    of three runs.}
  \label{fig:tpot-tput}
\end{figure}

Two further observations point away from the cache. First, the median and
the tail diverge. At 0.12 agent graphs per second the medians agree, yet the
95th percentiles differ by $8.10\times$ ($[2.0, 32.0]$). A uniformly slower
engine would move both percentiles; a gap confined to the tail means that
most requests are unaffected while a minority wait much longer, the signature
of intermittent blocking (Section~\ref{sec:variability} examines the full
distributions). Second, decode speed moves the other way. TPOT agrees within
6\% at the two lowest rates, but at 0.20 and 0.28 agent graphs per second
SGLang decodes 12\% and 24\% more slowly (Figure~\ref{fig:tpot-tput}). Lower
TTFT bought with higher TPOT is exactly the trade of chunked prefill
described in Section~\ref{sec:bg-chunked}, and SGLang is the engine that
chunks by default.

\subsection{Decomposing the gap}
\label{sec:res-decomp}

The matched arm changes one setting of one engine. SGLang's resolved
configuration is identical in both arms, so its measurements test the
apparatus itself. They reproduce closely but not exactly. The matched runs
executed on a separately provisioned cluster days after the defaults arm.
Across nine latency measures (median and 95th-percentile TTFT and median
end-to-end latency, from sending a request to receiving its last token, at
three conditions), SGLang's values agree with the
originals within 0.5 to 11.7\% (Table~\ref{tab:reproducibility}), and TPOT
agrees within 1.3\%. All nine differences point the same way: the second
deployment is faster. Because seed pairing removes workload variation, the
paired test resolves this shift, and three of the nine differences reach
$p < 0.05$. The shift is small and consistent, the signature of a change in
the deployment rather than in the engine. The changes in vLLM between the
arms are $3.35\times$ or larger, so a shift of this size cannot produce them
(Section~\ref{sec:threats} discusses the remainder).

\begin{table}[t]
  \caption{SGLang under an identical configuration in two independently
    provisioned deployments, geometric means of three seed-paired runs. E2E
    is end-to-end latency. $^*$\,paired $p < 0.05$.}
  \label{tab:reproducibility}
  \small
  \begin{tabular}{llrrr}
    \toprule
    Condition & Metric & First deployment & Second deployment & Change \\
    \midrule
    Rate 0.12 & TTFT p50 & 0.468\,s & 0.458\,s & $-2.2\%$ \\
    Rate 0.12 & TTFT p95 & 1.311\,s & 1.241\,s & $-5.4\%$ \\
    Rate 0.12 & E2E p50 & 13.75\,s & 13.65\,s & $-0.8\%$ \\
    Rate 0.28 & TTFT p50 & 1.089\,s & 0.961\,s & $-11.7\%^*$ \\
    Rate 0.28 & TTFT p95 & 17.80\,s & 16.72\,s & $-6.1\%$ \\
    Rate 0.28 & E2E p50 & 21.17\,s & 20.80\,s & $-1.8\%^*$ \\
    8,192 tokens & TTFT p50 & 0.891\,s & 0.866\,s & $-2.8\%$ \\
    8,192 tokens & TTFT p95 & 4.221\,s & 4.088\,s & $-3.2\%^*$ \\
    8,192 tokens & E2E p50 & 15.50\,s & 15.43\,s & $-0.5\%$ \\
    \bottomrule
  \end{tabular}
\end{table}

Enabling chunked prefill moves vLLM a long way. Its TTFT improves
$3.5\times$ at the 95th percentile at 0.12 agent graphs per second,
$3.35\times$ at the median at 0.28, and $16.8\times$ at the median for
8,192-token prompts. Its TPOT worsens in step, from 112 to 125\,ms at 0.28.
That is the direction Agrawal et al.\ predict~\cite{agrawal2024sarathi}, and
it confirms that the setting took effect.

Table~\ref{tab:decomposition} and Figure~\ref{fig:decomp} decompose the gap
with the removed fraction of Equation~\eqref{eq:removed}. On TTFT, matching
removes 30 to 67\% of the default gap. On throughput the effect is larger:
the capacity gap at 0.28 agent graphs per second falls from $1.22\times$ to
$1.046\times$, a removed fraction of 78\%. The two measures disagree most at
the same condition, where TTFT retains a $12.3\times$ gap after matching.
Section~\ref{sec:res-queue} explains the disagreement.

\begin{table}[t]
  \caption{Engine gap $\bar{R}$ (vLLM divided by SGLang for latency, SGLang
    divided by vLLM for throughput) under vendor defaults and under matched
    chunked prefill, with the removed fraction $\phi$ of
    Equation~\eqref{eq:removed}. No fraction is given where the default gap
    is below 10\%. $^*$\,paired $p < 0.05$.}
  \label{tab:decomposition}
  \small
  \begin{tabular}{llrrr}
    \toprule
    Condition & Metric & Defaults & Matched & Removed \\
    \midrule
    Rate 0.12 & TTFT p50 & 1.09 & 0.95$^*$ & --- \\
    Rate 0.12 & TTFT p95 & 8.10$^*$ & 2.44$^*$ & 57\% \\
    Rate 0.28 & TTFT p50 & 36.4$^*$ & 12.3$^*$ & 30\% \\
    Rate 0.28 & TTFT p95 & 4.16$^*$ & 1.70 & 63\% \\
    8,192 tokens & TTFT p50 & 64.1$^*$ & 3.94 & 67\% \\
    8,192 tokens & TTFT p95 & 27.0$^*$ & 6.43$^*$ & 43\% \\
    \midrule
    Rate 0.28 & Throughput & 1.22$^*$ & 1.046 & 78\% \\
    \bottomrule
  \end{tabular}
\end{table}

\subsection{Latency near saturation measures the queue}
\label{sec:res-queue}

At 0.28 agent graphs per second under matched configuration, the offered
load is 1.68 requests per second. SGLang completes 1.66 and keeps pace. vLLM
completes 1.58, 4.4\% fewer, and falls behind. Its backlog then
grows for as long as requests keep arriving. Throughput and TTFT describe the
same situation from opposite ends: throughput reports that SGLang serves 4.6\% more
requests per second, and TTFT reports the waiting time that the difference
accumulates.

A fluid queue model makes the relation quantitative. Both engines receive the
same trace, so where SGLang keeps pace its throughput $X_s$ estimates the
trace's realised arrival rate. If vLLM serves at $X_v < X_s$, its backlog
grows at $X_s - X_v$, and a request arriving at time $t$ waits about
$(X_s - X_v)\,t / X_v$. Writing $\delta = 1 - X_v / X_s$ for vLLM's
throughput shortfall, the request arriving halfway through an arrival window
of length $T$ waits
\begin{equation}
  w_{50} \approx \frac{\delta}{1 - \delta} \cdot \frac{T}{2} .
  \label{eq:fluid}
\end{equation}
We predict vLLM's median TTFT as SGLang's median TTFT in the same repetition
plus $w_{50}$. The model has no free parameter.

\begin{figure}[t]
  \centering
  \includegraphics{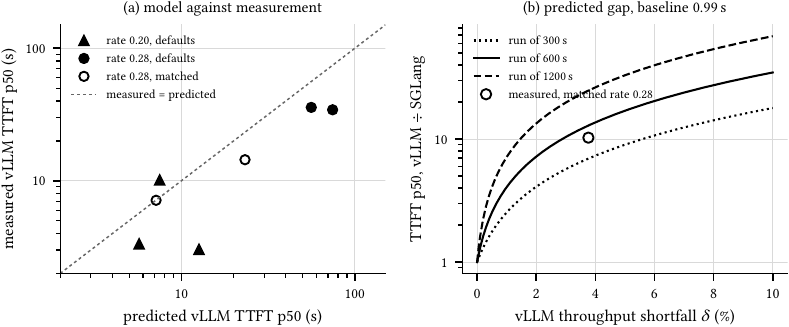}
  \caption{Fluid queue model of Equation~\eqref{eq:fluid}. (a)~Predicted
    against measured median TTFT of vLLM in the seven runs where vLLM falls
    behind and SGLang keeps pace. (b)~Predicted engine gap against vLLM's
    throughput shortfall for three run lengths, with SGLang's measured median
    as the baseline, and the measured gap at matched rate 0.28.}
  \label{fig:queue}
\end{figure}

Figure~\ref{fig:queue}a tests the model on the seven runs where vLLM falls
more than 1\% behind and SGLang keeps pace. Predictions and measurements
correlate with $r = 0.85$ on the log scale. The model overestimates, by a
factor of 1.6 on average, and the workload explains why. Workers are issued
only after the planner completes, and the aggregator only after the workers,
so a congested engine delays the arrival of its own later requests. A
partially closed system of this kind accumulates a smaller backlog than the
open fluid model assumes. At matched rate 0.28, in the two repetitions where
SGLang kept pace, the model predicts a median gap of $13.1\times$ from the
throughput shortfall alone; the measured gap is $10.3\times$. Below capacity
the model does not apply. There the apparent shortfall is 0.1 to 0.2\%, the
requests still in flight when a run ends, and no backlog forms.

Equation~\eqref{eq:fluid} has a consequence for any latency benchmark run
near saturation. Beyond the knee, the queueing delay is proportional to $T$
(Figure~\ref{fig:queue}b). The same 4.4\% shortfall yields a gap about half as
large in a 300\,s run and about twice as large in a 1,200\,s run. A TTFT
ratio measured there is a property of the engine and of the benchmark's
duration together, and it cannot be compared across studies without
throughput beside it.

\subsection{What survives matching}
\label{sec:res-tail}

Two conditions retain a significant gap that queueing cannot explain. At
0.12 agent graphs per second both engines keep pace. They complete 0.769 and
0.770 requests per second against 0.72 offered, so no backlog forms. Their medians are close, with vLLM marginally faster
($0.95\times$, $[0.94, 0.95]$), yet SGLang's 95th percentile remains
$2.44\times$ better ($[1.08, 5.53]$). With 8,192-token prompts the engines
complete essentially the same work, 0.345 and 0.349 requests per second, both
near 89\% of the offered load. Their queues therefore grow at nearly the same
rate, and still SGLang's 95th percentile is $6.43\times$ better
($[1.83, 22.55]$).

When capacity and saturation are matched, the order in which waiting
requests are served remains to separate the engines. SGLang admits first the
request with the longest cached prefix; vLLM serves in arrival order
(Section~\ref{sec:bg-order}). Favouring requests whose prefill is cheapest
would shorten the tail without changing throughput, which is the pattern we
observe. We state this as a hypothesis. Our experiments hold chunk size
constant but not request ordering, attention kernels or memory-budget
semantics (Table~\ref{tab:config}), and none of these is isolated
(Section~\ref{sec:threats}).

\subsection{Sensitivity to prefix sharing}
\label{sec:res-sigma}

\begin{figure}[t]
  \centering
  \includegraphics{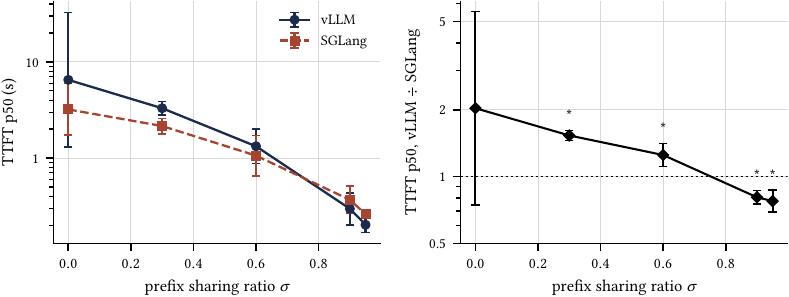}
  \caption{Left: median TTFT against prefix sharing ratio $\sigma$ under vendor
    defaults. Right: paired engine gap; $^*$\,paired $p < 0.05$. This sweep
    was not repeated under matched configuration.}
  \label{fig:sigma}
\end{figure}

If the gap came from the cache architecture, it would widen with $\sigma$,
since a finer-grained cache has more to gain when more of each prompt is
shared. The data show the opposite (Figure~\ref{fig:sigma}). The paired gap
falls from $1.53\times$ at $\sigma = 0.3$ to $0.81\times$ at $\sigma = 0.9$,
crossing parity between 0.6 and 0.9, and is significant at four of five
sharing levels. At $\sigma \ge 0.9$ vLLM is significantly faster. The result
at $\sigma = 0$ is inconclusive, with an interval of $[0.75, 5.54]$: one
repetition drew a burstier arrival realisation that affected both engines.

This sweep ran only under vendor defaults, and the chunked-prefill difference
predicts the same pattern. At low $\sigma$ each prompt is long and mostly
uncached, so vLLM's unchunked prefills block the queue. At high $\sigma$ most
of each prompt is served from cache, the remaining prefill is short, chunking
has little to do, and vLLM's lower scheduling overhead prevails. We
therefore report the reversal as an observation about default
configurations. Separating the two explanations needs a matched sharing
sweep.

\begin{figure}[t]
  \centering
  \includegraphics{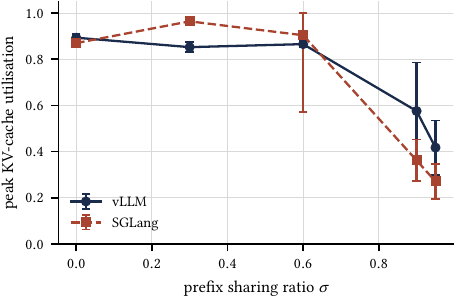}
  \caption{Peak KV-cache utilisation against prefix sharing ratio under vendor
    defaults. Arithmetic means of three runs.}
  \label{fig:kv}
\end{figure}

Prefix caching did engage in both engines. Peak KV-cache utilisation falls
from 0.89 to 0.42 for vLLM and from 0.87 to 0.27 for SGLang as $\sigma$ rises
from 0 to 0.95 (Figure~\ref{fig:kv}), consistent with each engine storing a
shared prefix once rather than once per request. No run preempted more than
three requests. We do not report per-run prefix-cache hit rates. Both engines
expose the hit rate as a counter accumulated since process start, and on the
sharing sweep, where runs of different $\sigma$ follow one another against one
engine process, the counter mixes runs. On the rate sweep, where every run
used $\sigma = 0.9$, it reads 0.896 for vLLM and 0.900 for SGLang, matching
the configured value.

\section{Performance Variability}
\label{sec:variability}

A median summarises a run, and it hides two things. One is how latency is
spread across the requests of a run. The other is how far a measurement
repeats when only the seed changes. This section examines both, then asks
which requests wait, how the choice of percentile shapes the picture, and
what restarting an engine costs.

\subsection{Distribution of request latency}
\label{sec:var-dist}

\begin{figure}[t]
  \centering
  \includegraphics{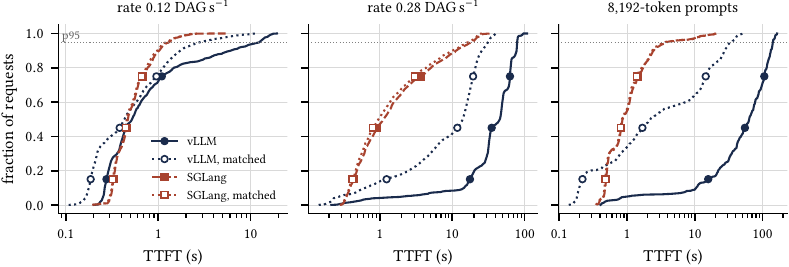}
  \caption{Empirical distribution of TTFT per request, pooled over three
    repetitions, with the 95th percentile marked. Line style and marker shape
    identify the engine, hollow markers the matched configuration.}
  \label{fig:ttft-dist}
\end{figure}

Figure~\ref{fig:ttft-dist} plots the empirical distribution of TTFT, pooled
over the three repetitions, for the three conditions measured under both
configurations. Table~\ref{tab:dispersion} summarises each distribution by its
tail ratio, the 95th percentile divided by the median, and by the skewness
$g_1 = \mathrm{E}[(Y - \mu)^3] / \sigma_Y^3$ of $Y = \ln \mathrm{TTFT}$. A
positive $g_1$ marks a long right tail over a fast majority. A negative $g_1$
marks mass concentrated near the top of the distribution.

\begin{table}[t]
  \caption{Dispersion of per-request TTFT, pooled over three repetitions. The
    tail ratio is the 95th percentile divided by the median, and $g_1$ is the
    skewness of $\ln$ TTFT. Generated by \texttt{make\_dispersion\_table.py}.}
  \label{tab:dispersion}
  \small
  \begin{tabular}{llrrrr}
    \toprule
    & & \multicolumn{2}{c}{vLLM} & \multicolumn{2}{c}{SGLang} \\
    \cmidrule(lr){3-4}\cmidrule(lr){5-6}
    Condition & Configuration & Tail ratio & $g_1$ & Tail ratio & $g_1$ \\
    \midrule
    Rate 0.12 & defaults & 24.0 & +1.29 & 2.8 & +0.93 \\
 & matched & 6.6 & +0.79 & 2.7 & +0.91 \\
\midrule
Rate 0.28 & defaults & 1.7 & $-$2.54 & 16.4 & +0.71 \\
 & matched & 2.2 & $-$1.16 & 17.7 & +0.80 \\
\midrule
8,192 tokens & defaults & 2.2 & $-$1.93 & 4.5 & +1.40 \\
 & matched & 15.4 & $-$0.08 & 4.4 & +1.44 \\
\bottomrule

  \end{tabular}
\end{table}

At 0.12 agent graphs per second under vendor defaults, vLLM serves most
requests quickly and a minority slowly. The median is 0.52\,s and the 95th
percentile 12.5\,s, a tail ratio of 24.0 with $g_1 = +1.29$. SGLang's ratio is
2.8. Matching cuts vLLM's ratio to 6.6, and SGLang's remains lower at 2.7.

At 0.28 agent graphs per second the signs reverse. Under vendor defaults,
vLLM has a tail ratio of 1.7 and $g_1 = -2.54$. The median request waits
45.5\,s. Because a queue that grows throughout the run delays almost every
request, even the 95th percentile reaches only 79\,s, and the distribution
piles up against its upper end. Matching moves the median to 13.9\,s and the
ratio to 2.2, which is still a saturated queue. SGLang's tail ratio is 16.4
under defaults and 17.7 matched, with positive skew in both, and it kept pace
with the offered load in two of three repetitions at this rate. Its tail ratio
peaks here, which places SGLang just below its own knee.

With 8,192-token prompts, vLLM under vendor defaults is saturated: tail ratio
2.2, $g_1 = -1.93$, median 63.3\,s. Under matched configuration its
distribution is the widest we measured. The ratio is 15.4 and $g_1$ is near
zero. The 75th percentile is 23.5 times the 25th, against at most 6.9 for any
other of the twelve distributions.

The two statistics separate two kinds of slowness. A large tail ratio with
positive skew marks intermittent blocking, the case of Section~\ref{sec:res-defaults}.
A ratio near two with negative skew marks a queue that has taken over the
whole distribution.

\subsection{Variability across repetitions}
\label{sec:var-spread}

\begin{figure}[t]
  \centering
  \includegraphics{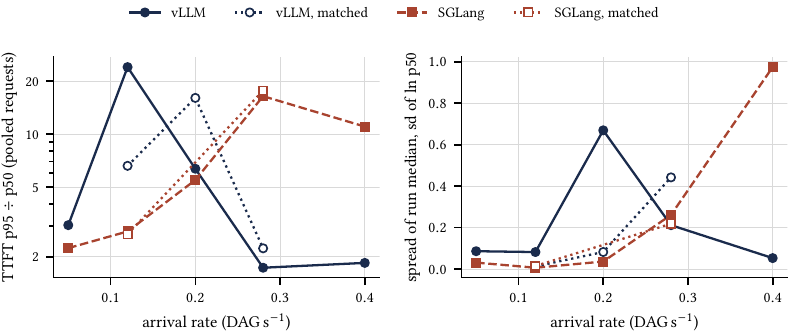}
  \caption{Variability against arrival rate. (a)~Tail ratio of pooled
    per-request TTFT, 95th percentile divided by median. (b)~Spread of the
    run-level median TTFT across the three repetitions, as the standard
    deviation of $\ln$ p50. Markers and lines as in
    Figure~\ref{fig:rate}.}
  \label{fig:spread}
\end{figure}

The three repetitions of a condition differ only in their seed. Figure~\ref{fig:spread}b
shows how far their median TTFT values disagree, as the standard deviation of
$\ln$ p50. Below the knee, at 0.12 agent graphs per second and lower, the
median spread is 0.01 for SGLang and 0.08 for vLLM, so repetitions agree within
about 1\% and 8\%. Above it, at 0.20 and higher, the median spread is 0.24 for SGLang and 0.21
for vLLM.

Spread peaks at each engine's knee. vLLM's reaches 0.67 at 0.20 agent graphs
per second, where its three medians are 3.0, 3.3 and 10.1\,s. SGLang's reaches
0.97 at 0.40, where its medians are 6.2, 29.3 and 4.8\,s. The prompt-length
sweep shows the same contrast for vLLM. At 2,048 tokens the spread is 0.07 and
the three medians lie between 0.25 and 0.29\,s. At 512 and 8,192 tokens it
is 1.06 and 0.61, with medians ranging from 0.7 to 6.1\,s and from 36 to
114\,s. Matched, the 8,192-token medians are 3.2, 9.8 and 1.3\,s, a spread of
1.03. It explains why the matched median gap in Table~\ref{tab:decomposition}
has an interval that includes parity. The tail ratio in
Figure~\ref{fig:spread}a peaks at the same place. For vLLM it peaks at 0.12 and
falls to 1.7 by 0.28 as the whole distribution moves up. For SGLang it peaks
at 0.28. Near the knee, a single burst of arrivals decides whether a queue
starts to grow.

Decode speed does not behave this way. The spread of median TPOT across
repetitions has a median of 0.021 over all conditions and never exceeds
0.09. A noisy platform would affect TPOT too, so this argues against
platform noise as the source of the variability in TTFT. The variability does
account for the width of the confidence intervals of Section~\ref{sec:results}, which are widest at the knee. The
interval on the 0.28 gap, for example, spans a factor of 6.9.

\subsection{Which requests wait}
\label{sec:var-roles}

Each agent graph contains three kinds of request, and they do not wait
equally. At 0.28 agent graphs per second, workers wait longer than planners
under every combination of engine and configuration. SGLang's median TTFT is
1.42\,s for workers and 0.54\,s for planners under vendor defaults, a factor
of 2.6, and 1.21\,s against 0.51\,s when matched. For vLLM the factors are 1.3
under vendor defaults (46.9\,s against 35.7\,s) and 1.7 when matched (15.0\,s
against 9.0\,s). Aggregators fall between the two. Each graph releases its four
workers within a short interval after the planner completes, so a worker
usually finds siblings queued ahead of it. The data are consistent with this
account, and we did not test it directly.

The vLLM tail at light load, in contrast, does not depend on the role. At 0.12
agent graphs per second under vendor defaults the tail ratio is 31 for
planners, 21 for workers and 23 for aggregators, against 2.0, 2.6 and 2.1 for
SGLang. A planner is the one request in a graph with no siblings, so it
reflects the engine's state at the moment it arrives. Its long tail shows that
a long prefill stalls whichever request happens to arrive during it.

\subsection{The median and the tail}
\label{sec:var-tail}

\begin{figure}[t]
  \centering
  \includegraphics{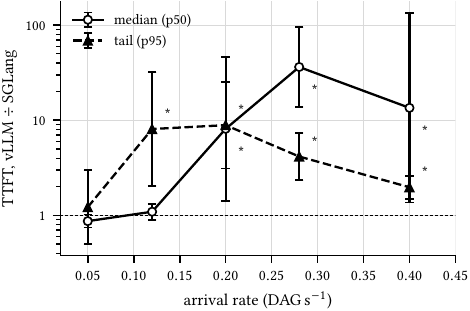}
  \caption{Paired engine gap in median and 95th-percentile TTFT across the
    rate sweep under vendor defaults. Values above 1 indicate that SGLang is
    faster. $^*$\,paired $p < 0.05$.}
  \label{fig:median-tail}
\end{figure}

Figure~\ref{fig:median-tail} compares the two paired gaps across the rate
sweep. The tail gap opens first. At 0.12 agent graphs per second it is
$8.10\times$ while the median gap is $1.09\times$. The two gaps are nearly equal at 0.20 ($8.13\times$ at the median and
$8.89\times$ at the 95th percentile), and the median gap dominates at 0.28,
where it reaches $36.4\times$ against $4.16\times$. At 0.40 the tail gap
closes to $1.98\times$ as both engines saturate and their whole
distributions move together.

The choice of percentile therefore moves the apparent onset of degradation.
A benchmark that reports the median finds vLLM's gap significant from 0.20
agent graphs per second, and one that reports the 95th percentile finds it
significant from 0.12, at a rate 1.7 times lower. We report both for this
reason, and Section~\ref{sec:discussion} returns to the point.

\subsection{Restarting an engine}
\label{sec:var-cold}

\begin{figure}[t]
  \centering
  \includegraphics{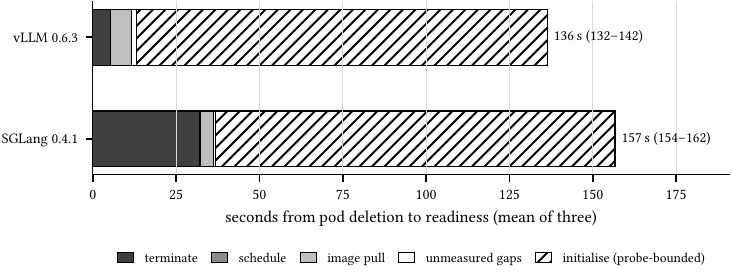}
  \caption{Time from pod deletion to readiness, decomposed by phase, mean of
    three repetitions. The GPU node stayed provisioned. Totals show the mean
    and the range. The initialisation phase is quantised by the readiness
    probe and is an upper bound.}
  \label{fig:cold}
\end{figure}

We deleted the serving pod and measured the time until its replacement became
ready, holding the GPU node provisioned so that engine initialisation is
separate from node provisioning (Section~\ref{sec:des-protocol}). The
restarts ran during the matched campaign, with each engine in its matched
configuration. Event timestamps have a resolution of one second.

From deletion to readiness, vLLM took 136\,s (range 132 to 142\,s) and SGLang
157\,s (154 to 162\,s), as Figure~\ref{fig:cold} shows. The difference comes
almost entirely from termination. SGLang's pod took 32\,s to shut down in
all three repetitions and vLLM's 5 or 6\,s. We did not investigate why.
Scheduling was instantaneous, and image pulls took 1 to 9\,s because the node
already held both images.

Initialisation, from container start to readiness, took exactly 120\,s in
five of the six runs. The sixth, vLLM's third, took 130\,s. The regularity
matches the readiness probe's schedule, with a 60\,s initial delay and a 10\,s
period, so we read 120\,s as the first probe at which the engine answered, not
as its loading time. It is an upper bound for both engines, and we draw no
comparison between them on this phase. SGLang's six-fold longer shutdown is
the one difference the data support, and it affects how quickly a replica
leaves service when a scaling policy removes it.

\section{Discussion}
\label{sec:discussion}

\subsection{What the decomposition shows}
\label{sec:disc-meaning}

The question was whether the performance gap between two prefix-caching
engines reflects their cache architectures. For the pair we measured, most of
it does not. One scheduling setting, chunked prefill, accounts for 78\% of the
capacity gap at the load where that gap is largest and for 30 to 67\% of the
latency gap across conditions. A comparison of default deployments cannot
separate the two causes, because the engines' defaults vary both at once.

A difference remains, and it is smaller and more specific. At equal
throughput SGLang's 95th percentile is better by 2.4 to $6.4\times$. That
pattern points to the order in which waiting requests are served. Whether the
cause is request ordering, an attention kernel or the engines' memory-budget
semantics is a question our experiments do not settle, and an ablation could.
SGLang exposes its scheduling policy as an option, so running it with
first-come-first-served ordering is a direct test.

Two observations carry beyond this pair of engines. The first concerns the
mechanism. On this workload, interference between prefill and decode explains
the larger part of the capacity gap and, in three of five conditions, of the
latency gap, and the layout of the cache explains less. Disaggregated designs
attack the same interference by placing the two phases on different
hardware~\cite{zhong2024distserve,patel2024splitwise}. The second concerns
measurement. Near saturation, a latency ratio is a property of the engine and
of the length of the benchmark together (Section~\ref{sec:res-queue}). A
ratio of $12\times$ can stand for a capacity difference under 5\%.

\subsection{Recommendations for comparing serving engines}
\label{sec:disc-rec}

Each recommendation follows from a result above.
\begin{enumerate}[leftmargin=1.6em,itemsep=2pt,topsep=4pt]
  \item \textbf{Record the resolved configuration.} Both engines choose their
        chunked-prefill setting from the hardware and the context length.
        Read it from the start-up log, not from the flags passed.
  \item \textbf{Change one setting at a time, and keep a control.} Matching
        chunked prefill altered one engine's one setting. SGLang, left
        unchanged, showed that independently provisioned deployments agree
        within 12\%.
  \item \textbf{Report throughput beside latency.} Say whether each engine
        keeps pace with the offered load. Without it, a latency ratio near the
        knee measures a queue.
  \item \textbf{State the run length when any engine is saturated.} The queueing
        delay grows in proportion to it.
  \item \textbf{Report the median and the 95th percentile.} The percentile
        moved the apparent onset of degradation by a factor of 1.7 in
        arrival rate (Section~\ref{sec:var-tail}).
  \item \textbf{Pair runs by seed.} The paired test found a significant
        difference at four sharing levels where the unpaired test found one.
\end{enumerate}

\subsection{Relation to prior claims}
\label{sec:disc-prior}

The RadixAttention authors report throughput up to $6.4\times$ higher than
contemporaneous systems~\cite{zheng2024sglang}. Their measurements used
different workloads, hardware and engine versions, so ours neither confirm
nor contradict that figure. Our result concerns a narrower matter: what a gap
between current default deployments can be attributed to.

Agrawal et al.\ diagnosed the generation stall and introduced chunked
prefill to remove it~\cite{agrawal2024sarathi}. We apply their diagnosis in a
new setting. The technique they proposed to resolve a throughput and latency
trade-off within one engine also accounts for much of the trade-off observed
between two engines that each adopted it. Our TPOT results reproduce the cost
they describe: vLLM's median TPOT at 0.28 agent graphs per second rose from 112 to
125\,ms once chunking was enabled.

Systems for multi-agent workloads add mechanisms above the cache. Examples
are eviction that follows the dependency graph~\cite{pan2025kvflow},
KV-cache scheduling aware of the workflow topology~\cite{bian2025tokencake},
prefix-aware request routing~\cite{srivatsa2025preble} and exposing request
structure to the serving layer~\cite{lin2024parrot}. These mechanisms act on
axes we did not vary, and our results do not bear on them. Where a design's
motivation rests on a comparison between default deployments of vLLM and
SGLang, the comparison varies cache and scheduling together, and the gap it
shows is an upper bound on the effect of the cache.

The pairing we measured has since changed upstream. vLLM's V1 engine enables
chunked prefill by default wherever possible~\cite{vllm2025v1guide}, so
current defaults of the two engines no longer differ in that respect. We read
this as support for the broader point. Defaults depend on hardware and
version, and they move. A comparison between current releases needs its own
record of resolved configurations, and the protocol of Section~\ref{sec:design}
applies to it unchanged.

\subsection{What this study adds}
\label{sec:disc-unique}

To our knowledge, this is the first controlled decomposition of the
vLLM and SGLang performance gap that separates chunked-prefill configuration
from cache architecture. It adds a result specific to the pair: the 78\% and
the 30 to 67\%. It adds a general result. Near saturation the latency gap
measures queue growth and scales with the length of the run, and a fluid model
with no free parameter reproduces it. It also adds an apparatus that others can
repeat, namely a testbed defined in code, a seeded workload generator, and the
per-run logs behind every number.

The decomposition is also falsifiable. A different GPU, context length or
release changes the confound, and the same protocol re-measures it. The
residual becomes a testable hypothesis, equal throughput with an unequal tail,
and Section~\ref{sec:conclusion} lists the experiments that would test it.

\section{Threats to Validity}
\label{sec:threats}

\subsection{Construct validity}
\label{sec:thr-construct}

\paragraph{TTFT includes queueing.} TTFT measures the wait for admission as
well as the prefill. This is deliberate, since a user experiences both, but it
means that TTFT near saturation reports the state of a queue
(Section~\ref{sec:res-queue}). We pair every latency with throughput for this
reason.

\paragraph{The keeps-pace criterion is a convention.} We call an engine
saturated when it completes less than 97\% of the offered load
(Section~\ref{sec:met-run}). At the condition that matters most, matched 0.28
agent graphs per second, SGLang achieves 98.5\% of the offered load and vLLM
94.2\%, so any tolerance between 1.5 and 5.8\% gives the same classification of
the two engines. The classification of individual repetitions near the
threshold can change, and the fluid model of Section~\ref{sec:res-queue} uses
it to select runs.

\paragraph{The workload is synthetic.} Prompts are token sequences sampled from
the vocabulary. Prefix-cache matching compares token identifiers, so content
does not affect it, and the generator controls $\sigma$ and prompt length
exactly. It omits the length distributions of real prompts and near-duplicate
variation such as reordered tool schemas, which lower hit rates in deployment.
Tool-call latencies are simulated draws from published distributions, and every
graph has the same wide-branch topology. We did not test deeper chains or a
mix of topologies. Production characterisations report that arrival
patterns shift over time and that no single stochastic process fits every
workload~\cite{xiang2025servegen}, so a stationary Gamma process with one
coefficient of variation is a simplification.

\subsection{Internal validity}
\label{sec:thr-internal}

\paragraph{The residual is not attributed.} Matching chunked prefill leaves
several differences between the engines: the order of waiting requests,
attention kernels, SGLang's overlapped scheduler, and memory-budget parameters
that are defined differently (Table~\ref{tab:config}). Section~\ref{sec:res-tail}
names request ordering as the likeliest cause, and we did not isolate it.

\paragraph{The explanation of the sharing sweep is post hoc.} We expected the
gap to grow with $\sigma$ and it did not. We formed the account we give, that the gap
tracks prefill scheduling and that sharing shrinks the work scheduling has
to do, after seeing the data. It agrees with the decomposition and
with the KV-cache measurements, and it needs independent confirmation. The
sweep itself ran only under vendor defaults.

\paragraph{Deployment is confounded with configuration for vLLM.} The defaults
and matched arms ran on independently provisioned clusters. Within the matched
arm, the node replacements of Section~\ref{sec:des-protocol} also put the two
engines on different physical L4 instances of the same machine type at the
matched rates. SGLang, unchanged between arms, bounds the drift between
deployments at 11.7\% for that engine, and the drift has a direction: all
nine measures of Table~\ref{tab:reproducibility} are lower in the second
deployment. SGLang's token-level timing is just as stable. At the two rates
measured in both arms, the median request's 99th-percentile inter-token
latency differs by at most 19\% between arms in every paired repetition for
SGLang, while enabling chunked prefill lowers vLLM's by a factor of 2.2 to
2.3. We have no direct bound for vLLM. If it shared SGLang's shift, up to
about 12\% of its improvement between arms would come from the deployment
rather than the configuration. The effects we attribute to its
configuration are $3.35\times$ or larger, far beyond that figure, so ordinary
drift cannot account for them, though it could account for part of a small
one.

\paragraph{The fluid model is approximate.} It over-predicts vLLM's median
TTFT by a factor of 1.6 on average, because agent graphs issue later requests
only after earlier ones complete, so a congested engine slows its own arrivals.
It rests on seven runs, applies only above capacity, and takes SGLang's
throughput as the realised arrival rate, which holds only where SGLang keeps
pace. We use it to show that a small shortfall can produce the gap we measured
and that the gap scales with run length. We do not use it as a predictor.

\paragraph{Cache state and the hit-rate counter.} We wait 30\,s between runs and
exclude the first agent graph, but we cannot prove that no cached state
survives between runs. Both engines report the prefix-cache hit rate as a
counter accumulated since process start, so we do not report it per run on the
sharing sweep. The evidence that prefix caching engaged is the fall in KV-cache
utilisation, which is indirect.

\paragraph{The platform is shared.} The engine runs in a virtual machine whose
host we do not control. The spread of median TPOT across repetitions is small,
with a median of 0.021 (Section~\ref{sec:var-spread}), which argues against
large interference, and we cannot exclude a small one.

\subsection{External validity}
\label{sec:thr-external}

\paragraph{One GPU, one model, one context length.} All results come from an
NVIDIA L4 with Llama-3.1-8B-Instruct and a 16,384-token context. Both engines'
chunked-prefill defaults depend on these, so the out-of-the-box comparison
would differ on other hardware or at other context lengths. The conditionality
is part of our finding and also bounds it.

\paragraph{Engine versions.} We measured vLLM~0.6.3 and SGLang~0.4.1. vLLM's V1
engine has since changed the default~\cite{vllm2025v1guide}, so the specific
pairing we measured does not arise between current releases. We make no claim
about later versions of either engine. The protocol applies to them.

\paragraph{One replica.} Each engine ran as a single replica with autoscaling
disabled. Multi-replica deployments add routing and scaling effects that we
did not measure. The restart experiment holds the GPU node provisioned, so it
measures engine initialisation and not node provisioning, and the readiness
probe quantises its initialisation phase (Section~\ref{sec:var-cold}).

\subsection{Statistical validity}
\label{sec:thr-stat}

\paragraph{Three repetitions.} Each condition has $n = 3$. Confidence
intervals use $t_{0.975,2} = 4.30$ and are wide, and the tests have low power.
We read a non-significant result as an absence of evidence. Small differences
can reach significance when the runs agree closely, as the 5\% median
difference at 0.12 agent graphs per second does, so we do not rely on
significance alone to judge an effect's size.

\paragraph{Multiple comparisons.} The analysis computes 64 paired tests across
four metrics and the conditions of the study, and 42 fall below $p = 0.05$. If
no difference existed, about three would do so by chance. We apply no
correction. Our conclusions rest on effect sizes that exceed their intervals
by large factors and on the consistency of the pattern across conditions,
not on any single $p$-value.

\section{Related Work}
\label{sec:related}

\paragraph{Serving engines and prefix reuse.} Iteration-level
scheduling~\cite{yu2022orca} lets an engine rebuild its batch at every
decoding step, and both engines we study inherit it. PagedAttention adds
block-structured KV-cache memory and automatic prefix caching over hashed
blocks~\cite{kwon2023vllm}. RadixAttention indexes cached tokens in a radix
tree and reuses a shared prefix of any length~\cite{zheng2024sglang}. These
two mechanisms, and the comparison between them, are the subject of this
paper.

\paragraph{Prefill and decode scheduling.} Agrawal et al.\ identify the
generation stall and remove it by splitting prefills into chunks interleaved
with decode steps on the same GPU~\cite{agrawal2024sarathi}. Both engines we
study adopted this technique. DistServe~\cite{zhong2024distserve} and
Splitwise~\cite{patel2024splitwise} instead place prefill and decode on
separate hardware. Our result concerns the first approach: the presence or
absence of chunking is the configuration difference we isolate.

\paragraph{Serving multi-agent workloads.} Several systems target agent and
structured workloads directly. KVFlow manages prefix caching for multi-agent
workflows~\cite{pan2025kvflow}. TokenCake is a KV-cache-centric serving
framework for multi-agent applications~\cite{bian2025tokencake}. Preble
schedules prompts across a distributed deployment~\cite{srivatsa2025preble},
Parrot exposes application structure to the serving layer through semantic
variables~\cite{lin2024parrot}, and SAGA schedules agent workflows atomically
on GPU clusters~\cite{guo2026saga}. These systems change mechanisms that we
hold fixed. Section~\ref{sec:disc-prior} discusses how our results bear on
comparisons of this kind.

\paragraph{Workload and cost characterisation.} BurstGPT provides traces from
production LLM serving~\cite{wang2024burstgpt}, and its finding of bursty
arrivals informs our arrival layer. ServeGen characterises production
serving workloads across language, multimodal and reasoning models, and
reports that workloads generated naively are easier to serve than real
ones~\cite{xiang2025servegen}. Patil estimates the cost of LLM
infrastructure with a methodology that accounts for
concurrency~\cite{patil2026costperf}. These studies characterise a workload or a
single deployment. Neither compares engines under controlled configuration.

\paragraph{Performance-evaluation methodology.} Our methods draw on an
established literature. Fleming and Wallace show that benchmark ratios should
be summarised by their geometric mean~\cite{fleming1986lie}. Mytkowicz et
al.\ demonstrate that a seemingly innocuous aspect of an experimental setup can
lead to wrong conclusions~\cite{mytkowicz2009wrong}, and the same authors
analyse how omitted variables bias the evaluation of compiler
optimisations~\cite{mytkowicz2010omitted}. The chunked-prefill default is an
omitted variable of exactly this kind, set by the engine rather than chosen by
the experimenter. Hoefler and Belli propose twelve rules for reporting
performance results, after finding the state of practice wanting in a sample
of papers from three conferences~\cite{hoefler2015benchmarking}. Papadopoulos
et al.\ propose eight principles for reproducible performance evaluation in
cloud environments~\cite{papadopoulos2021principles}, and Iosup et al.\
measure how far the performance of production cloud services
varies~\cite{iosup2011variability}. Ilyushkin et al.\ compare seven autoscaling
policies experimentally on workflow workloads and find that parametrisation
matters a great deal~\cite{ilyushkin2018autoscalers}. Our study follows their
structure and applies the same discipline to LLM serving engines.

To our knowledge, no prior study isolates a configuration default as the
cause of a performance gap between LLM serving engines, or shows that a
latency gap measured near saturation scales with the length of the run.

\section{Conclusion and Future Work}
\label{sec:conclusion}

A large performance gap between two LLM serving engines, measured under
their default configurations, is usually read as a verdict on their cache
architectures. We tested that reading in a controlled experiment. Under
vendor defaults vLLM's median TTFT is up to $64\times$ SGLang's, but the
engines' defaults differ in whether they chunk long prefills. Matching that
one setting removes 78\% of the capacity gap at the load where it is largest
and 30 to 67\% of the latency gap. Near saturation, the latency gap that
remains measures the growth of a queue: a 4.4\% shortfall in throughput
produces a $12\times$ difference in median TTFT, and a fluid queue model
predicts that the difference grows with the length of the run. What survives
matching is a tail advantage of 2.4 to $6.4\times$ at equal throughput.

Several experiments would extend these results.
\begin{itemize}[leftmargin=1.4em,itemsep=2pt,topsep=4pt]
  \item \textbf{Isolating the residual.} Running SGLang with
        first-come-first-served ordering would test whether request ordering
        accounts for the tail advantage. Matching the memory budget and the
        attention kernel would address the other remaining differences.
  \item \textbf{A matched sharing sweep.} Repeating the sharing sweep with
        chunked prefill matched would separate the two explanations of the
        reversal in Section~\ref{sec:res-sigma}.
  \item \textbf{Current releases.} vLLM's V1 engine enables chunked prefill
        by default. Measuring current releases of both engines, with their
        resolved configurations recorded, would show which default
        differences remain.
  \item \textbf{Other hardware, models and workloads.} The defaults depend on
        GPU memory and context length, so other GPUs, larger models and
        traces from natural-language agent workloads would test how far the
        decomposition generalises.
  \item \textbf{A closed queueing model.} Agent graphs throttle their own
        arrivals when the engine falls behind. A model that captures this
        dependency should remove most of the fluid model's overestimate.
  \item \textbf{Scaling.} Multi-replica deployments with autoscaling, and
        restarts that include node provisioning, remain unmeasured.
\end{itemize}

\begin{acks}
We thank the Mitacs Globalink Research Internship program and Thompson
Rivers University for supporting this research.
\end{acks}

\bibliographystyle{ACM-Reference-Format}
\bibliography{references}

\appendix
\section{Artifact}
\label{app:artifact}

\subsection{Abstract}

The artifact contains everything needed to repeat the measurements and the
analysis: Terraform configuration for the testbed, Kubernetes manifests for
both engines in both configurations, the workload generator, the campaign
runners, the analysis and figure scripts, and the per-run logs from which
every table and figure is derived.

\subsection{Scope}

The artifact supports two levels of reproduction. \emph{Re-analysis}
regenerates every table and figure from the included logs in a few minutes,
without a GPU. It checks the statistics and the decomposition.
\emph{Re-measurement} provisions the testbed and repeats the experiments. The
full campaign takes about 33 hours of measured load on one GPU, 60\% of it the
sharing sweep. The central claim, the decomposition of
Table~\ref{tab:decomposition}, rests only on the three conditions measured in
both configurations, which take about 8 hours.

\subsection{Contents}

\begin{description}[leftmargin=1em,itemsep=2pt,topsep=4pt,font=\normalfont\ttfamily]
  \item[infra/] Terraform for the testbed in two stages, cluster and platform
        (Section~\ref{sec:sys-arch}), and the engine manifests of both
        configurations.
  \item[wlgen/] Trace compiler, validator, dispatcher, and the generators of
        the sweep and restart Jobs (Section~\ref{sec:sys-workload}).
  \item[runners/] The sweep and restart runners, which share one success
        check, the per-engine campaign script that ran the matched arm
        (Section~\ref{sec:des-protocol}), and a script that stages them for
        either configuration.
  \item[analysis/] Log parser, statistical analysis, and a script that
        recomputes every number quoted in this paper.
  \item[figures/] Figure and table generators and the shared figure style.
  \item[data/] Per-run logs and Prometheus extracts for all 99 runs, the
        engines' resolved start-up configurations, and the restart records.
        Logs of failed attempts and of the repetitions discarded after a node
        replacement are kept separately, each with a note giving the reason.
\end{description}

\subsection{Hosting}

The artifact is available at
\url{https://github.com/FaizanAnwar01/llm-serving-config-eval}. The commit used for this
paper is tagged.

\subsection{Hardware dependencies}

The testbed needs one GPU node with an NVIDIA L4 (24\,GB), for which we used
a GKE \texttt{g2-standard-8} instance, and one CPU node for the platform and
the generator, an \texttt{e2-standard-4} instance. It also needs a 60\,GB
persistent volume and GPU quota for one L4. The GPU matters. SGLang resolves
its default chunk size from GPU memory (Section~\ref{sec:sys-configs}), so a
different GPU compares a different pair of configurations.

\subsection{Software dependencies}

Google Kubernetes Engine (Kubernetes 1.35) and Terraform, with KServe,
kube-prometheus-stack and cert-manager installed by the second Terraform
stage. The same stage installs KEDA, which these experiments do not use. The model, Llama-3.1-8B-Instruct, is gated; access requires accepting
its licence and supplying a Hugging Face token to the cluster as a Kubernetes
Secret. The analysis needs Python~3.12 with the versions of pandas, NumPy,
SciPy and matplotlib pinned in the repository; regenerating the figures also
needs a LaTeX installation.
Both engines run from published images without modification
(Table~\ref{tab:digests}). A digest identifies an image exactly, whereas a
tag can be reassigned.

\begin{table}[h]
  \caption{Engine images. Digests are those of the published tags for
    linux/amd64, taken from the Docker Hub image pages. We did not record
    digests at run time.}
  \label{tab:digests}
  \footnotesize
  \begin{tabular}{ll}
    \toprule
    Image tag & Manifest digest \\
    \midrule
    \texttt{vllm/vllm-openai:v0.6.3} &
      \texttt{sha256:b8374cee0a1acaec8b64525ff77560f3} \\
      & \texttt{\ \ \ \ \ \ \ 0443f67bd0fc1956a3529504a89f823b} \\
    \texttt{lmsysorg/sglang:v0.4.1-cu121} &
      \texttt{sha256:50e7c37bda650334e505eb776e97ea92} \\
      & \texttt{\ \ \ \ \ \ \ 0de8dda4fd94811a877330fa684a5a37} \\
    \bottomrule
  \end{tabular}
\end{table}

\subsection{Workflow}

\begin{enumerate}[leftmargin=1.6em,itemsep=2pt,topsep=4pt]
  \item Apply the two Terraform stages, cluster and then platform.
  \item Deploy one engine in the chosen configuration, and confirm the
        resolved chunked-prefill setting in its start-up log before measuring.
  \item Generate the sweep with the paper's seed series and run it; each run
        compiles, validates and dispatches its trace as a Kubernetes Job and
        writes its results to a log. Repeat for the other engine.
  \item Extract each run's Prometheus window before tearing down the cluster,
        since the series cannot be recovered afterwards.
  \item Parse the logs, run the analysis, and check every quoted number
        with \texttt{make}; regenerate the figures and tables with
        \texttt{make figures}. The parser reports how many logs it found,
        parsed and skipped, and a non-zero skip count should be investigated
        before any output is trusted. The analysis refuses to pool runs from
        the two configurations.
  \item Destroy both Terraform stages to stop billing.
\end{enumerate}

\subsection{Expected results}

Re-analysis of the included logs reproduces every table and figure exactly,
because the analysis is deterministic; \texttt{make verify} confirms that the
regenerated outputs are byte-identical to the published ones. Re-measurement should reproduce each
engine's latencies within the variation we observed between our own
independent deployments, up to 12\% for SGLang under an unchanged
configuration (Table~\ref{tab:reproducibility}), and the engine ratios of the
decomposition within their confidence intervals. Absolute values will differ
on other GPUs and at other context lengths, because both engines' chunked
prefill defaults depend on them.

\end{document}